\documentclass[11pt]{article}
\usepackage[letterpaper,margin=1in]{geometry}
\usepackage[T1]{fontenc}
\usepackage{lmodern,amsmath,amssymb,mathtools}
\usepackage{graphicx}
\usepackage[labelsep=period,font=small]{caption}

\graphicspath{{figures/}}
\usepackage{placeins}

\usepackage{etoolbox}
\makeatletter\patchcmd{\l@section}{1.0em}{0.5em}{}{}\makeatother
\usepackage[nottoc]{tocbibind}
\usepackage[title,titletoc]{appendix}
\usepackage{xcolor}
\usepackage[numbers,sort&compress]{natbib}
\usepackage[colorlinks=true,linkcolor=blue!50!black,citecolor=blue!50!black,urlcolor=blue!45!black]{hyperref}
\newcommand{\tavg}[1]{\left\langle #1 \right\rangle}
\DeclareMathOperator{\tr}{tr}
\DeclareMathOperator{\re}{Re}
\DeclareMathOperator{\im}{Im}
\newcommand{\order}[1]{\mathcal{O}(#1)}

\title{Lyapunov spectrum of random neural networks}
\author{David G. Clark\\[0.5ex]
\small Center for Computational Neuroscience, Flatiron Institute, New York, NY 10010, USA\\
\small \href{mailto:dclark@flatironinstitute.org}{dclark@flatironinstitute.org}}
\date{October 8, 2026}

\begin{document}
\maketitle
{\renewcommand{\thefootnote}{}\footnotetext{Code and data: \url{https://github.com/davidclark1/rnn-lyapunov}.}}

\begin{abstract}
The Lyapunov spectrum of a nonlinear recurrent neural network with random asymmetric couplings is calculated in the
limit $N\to\infty$. The calculation is based on a finite-$N$ identity that expresses the
cumulative distribution of Lyapunov exponents as a response function of the tangent-space dynamics, with the tangent
trajectory selected by a minimum-norm condition rather than by an initial condition. A cavity method then
determines this response function at large $N$ through a self-consistent single-site problem. This result establishes
that the chaos in this network is extensive and gives access to diffeomorphism-invariant properties of the
dynamics. This work was done in collaboration with the AI models GPT-6 Astra and
Claude Opus 5.5.
\end{abstract}

\vspace*{\fill}
\pagebreak

{\small\tableofcontents}
\clearpage

\section{Introduction}
\label{sec:intro}

A defining feature of chaotic dynamical systems is extreme sensitivity to initial conditions, colloquially called the
\emph{butterfly effect}. In such systems, nearby trajectories separate exponentially in time. The rate of separation,
called a Lyapunov exponent, depends on the direction of the initial displacement, and the full set of rates defines the
Lyapunov spectrum. A generic displacement has a nonzero component along the most rapidly expanding direction, and thus
at late times the separation is governed by the maximum Lyapunov exponent. The full Lyapunov spectrum, however,
contains much more information than the maximum Lyapunov exponent alone, providing a rich characterization of
chaos. A canonical example of chaos in \emph{high} dimensions, described in Sec.~\ref{sec:rnn}, is the random neural network of
\citet{sompolinsky1988chaos}, who named the ``evaluation of the whole distribution of Lyapunov exponents'' of this
network as an open problem nearly four decades ago. The present work solves this problem.

\subsection{Lyapunov exponents}
\label{sec:lyap}

We consider dynamics in discrete time, $x(n+1)=G(x(n))$ with $x(n)\in\mathbb R^N$. This includes the Euler
discretization with time step $\delta$ of a continuous-time system $\dot x(t)=f(x(t))$, for which
$G(x)=x+\delta f(x)$. An infinitesimal perturbation of a trajectory, or tangent vector, $v(n)$ evolves under the
linearized (tangent-space) dynamics
\begin{equation}
    v(n+1)=M(n)\,v(n),
\end{equation}
where $M_{ij}(n)=\left.\partial G_i(x)/\partial x_j\right|_{x=x(n)}$ is the Jacobian of $G(x)$ evaluated at $x(n)$.
After $m$ steps, the perturbation is $v(m+1)=M(m)\cdots M(1)\,v(1)$. This product of Jacobians along a trajectory is
called the derivative cocycle. If $\sigma_1(m)\geq\cdots\geq\sigma_N(m)$ are the singular values of the derivative
cocycle, the Lyapunov exponents are
\begin{equation}
    \lambda_i=\lim_{m\to\infty}\frac{1}{m\delta}\log\sigma_i(m),\qquad i=1,\ldots,N.
    \label{eq:lyap}
\end{equation}
Since $m\delta$ is the elapsed physical time, the exponents $\lambda_i$ are rates per unit time rather than per step,
which makes the exponents $\lambda_i$ comparable across time steps. Additionally, as $\delta\to0$, the exponents
$\lambda_i$ approach those of the continuous-time system. Geometrically, the linearized dynamics deform an
infinitesimal ball of initial conditions into an ellipsoid whose principal axes grow or shrink exponentially at the
rates $\lambda_i$.

The existence of the Lyapunov exponents is established relative to an invariant measure, that is, a distribution of
states preserved by the dynamics. For a dissipative system, whose trajectories approach an attractor of zero volume, the relevant invariant measure is
the distribution of states visited by the trajectory of a typical initial condition in phase
space~\cite{young2002srb}. As in numerical studies, we assume that the dynamics have a single such measure and that
this measure is ergodic. For an ergodic invariant measure, the multiplicative ergodic theorem of Oseledets guarantees
that, under mild conditions on the Jacobians, the long-time limit in Eq.~\eqref{eq:lyap} exists and takes the same values for all
trajectories that are typical with respect to this measure~\cite{oseledets1968multiplicative}.

\subsection{Attractor dimension and entropy rate}
\label{sec:dimension_entropy}

The Lyapunov spectrum provides access to two basic properties of the dynamics, namely, the attractor dimension and the
entropy rate, which we now review. Since the Lyapunov spectrum is invariant under diffeomorphisms applied to the phase
space, these and other quantities derived from the Lyapunov spectrum are intrinsic properties of the dynamics,
independent of the arbitrary choice of coordinates.

We start with the attractor dimension. Note that the sum $\lambda_1+\cdots+\lambda_k$ is the growth rate of $k$-dimensional
volumes in tangent space. In a dissipative system, this rate is negative for $k=N$, since phase-space volume
contracts onto the attractor. If the system is also chaotic, that is, if $\lambda_1>0$, $k$-dimensional volumes grow
for sufficiently small $k$ and shrink for $k=N$. Thus, a
natural estimate of the attractor dimension is the value of $k$ at which growth and contraction balance. To this end, let $j$ be the
largest integer $k$ for which $\lambda_1+\cdots+\lambda_k\geq0$. Interpolating the sum $\lambda_1+\cdots+\lambda_k$
linearly between $k=j$ and $k=j+1$ and locating the zero of the interpolation gives the Kaplan--Yorke
dimension~\cite{KaplanYorke1979,frederickson1983liapunov}
\begin{equation}
    D_{\mathrm{KY}}=j+\frac{\lambda_1+\cdots+\lambda_j}{|\lambda_{j+1}|}.
    \label{eq:ky}
\end{equation}
Kaplan and Yorke conjectured that $D_{\mathrm{KY}}$ equals the information dimension of the invariant measure, which describes how the attractor fills phase space at increasingly fine resolution. This
conjecture has been proven in special cases~\cite{young1982dimension,ledrappier1985metric2}. In any case, $D_{\mathrm{KY}}$ is a
coordinate-invariant probe of the attractor dimension, and we refer to $D_{\mathrm{KY}}$ simply as the attractor
dimension.

We now turn to the entropy rate. Suppose we partition phase space into cells and record the sequence of cells that a
trajectory visits. Two initial conditions in the same cell by definition cannot be distinguished at the resolution of
the partition. Over time, for a generic displacement between the two initial conditions, the separation of the two
trajectories grows, so the two trajectories eventually visit different cells. The Kolmogorov--Sinai entropy rate
$h_{\mathrm{KS}}$ is the growth rate, per unit time, of the entropy of these symbolic sequences, maximized over
partitions~\cite{kolmogorov1958new,sinai1959notion}. That is, $h_{\mathrm{KS}}$ is the rate at which the dynamics reveal information about the initial condition.
According to Ruelle's inequality~\cite{ruelle1978inequality},
\begin{equation}
    h_{\mathrm{KS}}\leq\sum_{\lambda_i>0}\lambda_i.
    \label{eq:ruelle}
\end{equation}
Pesin's formula states that, under suitable conditions on the invariant measure~\cite{ledrappier1985metric1}, this inequality is an equality, i.e., $h_{\mathrm{KS}}$ equals the sum of the positive Lyapunov
exponents~\cite{pesin1977characteristic}. Pesin's formula thus plays the same role for the entropy rate that the Kaplan--Yorke conjecture
plays for the attractor dimension, linking a quantity defined through the invariant measure to a function of the
Lyapunov spectrum. In any case, the sum of the positive exponents is a coordinate-invariant estimate of the
entropy rate. In this paper, we write $h_{\mathrm{KS}}$ for the sum of the positive exponents and refer to
$h_{\mathrm{KS}}$ as the entropy rate.

\subsection{Random neural networks}
\label{sec:rnn}

The discussion so far applies to any smooth dynamical system. We now turn to the random neural network of
\citet{sompolinsky1988chaos}. This model consists of $N$ randomly coupled neurons, each of which low-pass filters its input
and applies a nonlinearity that models the conversion from membrane potential to firing rate. Networks of such neurons
are called rate networks. The chaotic phase of this
model has long served as a model of the spontaneous activity observed in cortex \emph{in vivo}. The model has also
been broadly influential in neuroscience~\cite{mastrogiuseppe2018linking}, particularly in light of the ability to train such networks to perform tasks~\cite{sussillo2009generating,driscoll2024flexible}.

As $N\to\infty$, dynamical mean-field theory describes the statistics of single-neuron activity in this model and
predicts a second-order transition to chaos as the coupling strength $g$ crosses the critical value
$g=1$~\cite{sompolinsky1988chaos,crisanti2018path}. Furthermore, \citet{sompolinsky1988chaos} obtained a
Schr\"odinger-type equation whose ground-state energy determines the maximum Lyapunov exponent. This approach has
since yielded the maximum Lyapunov exponents of networks with several populations of
neurons~\cite{kadmon2015transition} and of networks driven by external
input~\cite{schuecker2018optimal,engelken2022input}. At time
step $\delta=1$, where the network has no leak, the equations simplify and provide the maximum Lyapunov exponent in closed
form~\cite{molgedey1992suppressing,cessac1995increase}. 

In foundational prior work, \citet{engelken2023lyapunov} computed the full Lyapunov spectrum of this model numerically by
evolving a set of vectors under the tangent dynamics and periodically orthonormalizing them via a QR
decomposition~\cite{benettin1980lyapunov}, which keeps the vectors from growing without bound along the expanding
directions. When the exponents are plotted against their rank divided by $N$, the Lyapunov spectrum converges to a
limit as $N\to\infty$, indicating that the attractor dimension and the entropy rate are extensive, i.e.,
$D_{\mathrm{KY}}/N$ and $h_{\mathrm{KS}}/N$ converge to self-averaging values of order one. The linear scaling of these
quantities with $N$ is known as extensive chaos~\cite{ruelle1982large}, which has also been observed numerically in
discrete-time versions of this model~\cite{curato2013onset}. Both $D_{\mathrm{KY}}/N$ and $h_{\mathrm{KS}}/N$ vanish at
the onset of chaos and grow as the coupling strength increases beyond the onset. In continuous time ($\delta\to0$), the
attractor dimension remains below about one tenth of $N$ even at large $g$, so that the attractor, while extensive, is low-dimensional
relative to the full phase space.

Analytically, rather than numerically, the full Lyapunov spectrum has been obtained only in limiting regimes. In the stable phase, the
dynamics are linear around the trivial fixed point, at which all neurons are silent, and the Lyapunov spectrum follows
from the eigenvalue density of the coupling matrix, given by the circular
law~\cite{ginibre1965statistical,girko1985circular,sommers1988spectrum,curato2013onset}. In other limits,
\citet{engelken2023lyapunov} approximated the full Lyapunov spectrum by combining random-matrix theory with ans\"atze
for the Jacobians.

\citet{sompolinsky1988chaos}
argued that ``the number of positive Lyapunov exponents for [$g > 1$] is macroscopic, i.e., proportional to $N$,'' because nearby
trajectories diverge exponentially in the mean-field theory, which already describes the limit $N\to\infty$. This paper derives
the full Lyapunov spectrum of this model at large $N$, establishing as a byproduct that the chaos is extensive.

\subsection{Roadmap}
\label{sec:roadmap}

The derivation (Sec.~\ref{sec:derivation}) has three parts.

\begin{itemize}
    \item In the first part (Sec.~\ref{sec:part1}), which is exact at finite $N$, we express the cumulative distribution
    of Lyapunov exponents as a response function of the tangent dynamics. In particular, we first shift the tangent
    dynamics so that every exponent is lowered by a constant $s$; the cumulative distribution at $s$ is then the
    fraction of decaying tangent directions. We show that the one-step response of the minimum-norm solution of the
    shifted tangent dynamics driven by a source is the projector onto the decaying directions. Thus, the trace of this
    response gives the number of decaying directions, i.e., the number of Lyapunov exponents less than $s$.
    \item In the second part (Sec.~\ref{sec:dyn}), we cast the minimum-norm solution of the shifted tangent dynamics as
    the limit of the solution of a regularized least-squares problem as the regularization vanishes, and we write this problem as a forward--backward dynamical system
    with a $2N$-dimensional state, consisting of an $N$-dimensional field that runs forward in time and another that
    runs backward. The restriction of the solution of this system to the forward field solves the regularized problem.
    \item In the third part (Sec.~\ref{sec:part3}), we compute the forward-system response at large $N$ with the cavity method,
    applied jointly to this forward--backward system and the network dynamics, at fixed regularization. This involves
    adding one neuron to the network, along with its forward and backward fields. The effect of the rest of the network
    on this neuron reduces to the standard mean-field description of its activity and to two kernels for its forward
    and backward fields, which are determined self-consistently. The result is a single-site problem in which the
    time-varying gains of the neuron, that is, the slopes of its nonlinearity, are sampled from the known dynamical mean-field
    theory and drive the forward and backward fields. The solution of this single-site problem, with the regularization
    taken to zero at the end, gives the cumulative distribution of Lyapunov exponents at large $N$, assuming that the
    limit of large $N$ and the limit of vanishing regularization can be exchanged.
\end{itemize}

Following the derivation, in Sec.~\ref{sec:results}, we first show analytically that the single-site theory
recovers known results, and their straightforward extensions to finite $\delta$, for the Lyapunov spectrum at the
trivial fixed point and the maximum Lyapunov exponent in the chaotic phase. We then solve the single-site problem numerically and
compare the resulting Lyapunov spectra, attractor dimensions, and entropy rates with simulations of large networks.

\subsection{Related work}
\label{sec:related}

Prior analytical work on the dimension of the activity in this model has focused not on the attractor dimension but
on the so-called participation-ratio dimension, which provides a soft count of the dimensions of the linear subspace
explored by the activity. This subspace is identified by principal component analysis~\cite{litwin2017optimal}.
\citet{clark2023dimension} computed the participation ratio with a two-site cavity method, and
\citet{clark2025connectivity} later computed it from fluctuations around the saddle point of the path integral. These
calculations showed analytically
that the participation ratio is extensive. Because the participation ratio is a dimension of a linear subspace, it
depends on the choice of variables~\cite{jazayeri2021interpreting}, unlike the attractor dimension.

The maximum Lyapunov exponent and the participation ratio each reduce a spectrum---the Lyapunov spectrum and the
spectrum of the covariance matrix of the activity, respectively---to a single number. For the covariance matrix, the
full spectrum is now known from a recent line of work. For randomly coupled linear networks driven by noise that is
independent across neurons, the covariance matrix is an explicit function of the couplings, and the spectrum of the
covariance matrix follows from random-matrix theory~\cite{hu2022spectrum,zavatone2026stationary}. The covariance matrix
of the present nonlinear, autonomous network was then proposed~\cite{shen2025covariance} and analytically
shown~\cite{wakhloo2026solution,clark2026linear} to coincide with that of a linear network with the same couplings
driven by noise, an example of \emph{linear equivalence}. Importantly, this equivalence holds only for the covariance
matrix, and other properties of the two networks differ. In particular, the linear network is stable, with a negative
maximum Lyapunov exponent. The full Lyapunov spectrum must therefore be computed from the nonlinear dynamics directly,
as we do here.

Beyond rate networks, Lyapunov spectra have been computed for networks of spiking
neurons~\cite{monteforte2010dynamical,monteforte2012dynamic,lajoie2013chaos,lajoie2014structured,engelken2024sparse}.
Although spikes are discontinuous events, the state of the network, typically given by the membrane potentials of all neurons,
can be propagated exactly from one spike time to the next, defining a discrete-time map. The Lyapunov spectrum can then be
computed using products of the Jacobians of this map~\cite{monteforte2010dynamical}. In balanced networks (in which
large excitatory and inhibitory input currents approximately cancel, leaving strong fluctuations), such numerical
computations revealed extensive chaos~\cite{monteforte2010dynamical,lajoie2014structured}. These computations also
revealed intriguing phenomena such as ``stable chaos,'' in which the activity is irregular and unstable to finite
perturbations, such as the removal of a single spike, despite being stable to infinitesimal
perturbations~\cite{zillmer2006desynchronization,monteforte2012dynamic}.

Outside neural networks, \citet{ruelle1982large} argued that chaos should be extensive in spatially extended systems
comprising many weakly interacting subsystems. Systems with all-to-all coupling, such as the present
network, lack this structure, so extensive chaos in such systems is
nontrivial~\cite{takeuchi2009lyapunov,engelken2023lyapunov}.

In machine learning, signals propagating through deep networks at initialization undergo a transition between an
ordered and a chaotic phase~\cite{poole2016exponential}, and the proximity to this transition predicts how well
networks can be trained~\cite{schoenholz2017deep,cowsik2025geometric}. Relatedly, in recurrent networks, pushing the
Lyapunov exponents toward zero during training improves the learning of long-time dependencies~\cite{engelken2023gradient}.
Lyapunov analysis has also proved useful for interpreting the computations of trained networks. In modern \emph{looped} models that
reason by iterating a latent state toward a fixed point encoding the solution, the reasoning dynamics have been
suggested to become transiently chaotic on harder tasks, as trajectories linger near unstable states that correspond to nearly correct
solutions~\cite{lai2026fractal}. Additionally, in large language models, the Jacobian of the final-layer activations with respect to
intermediate activations defines directions in activation space that correspond to words the
model is disposed to produce, aiding interpretability~\cite{gurnee2026verbalizable}. A dynamical-systems description of how perturbations
propagate through a network, such as the one provided by the Lyapunov spectrum, may refine such analyses.

Methodologically, our calculation rests on the relation, derived in Part~1 (Sec.~\ref{sec:part1}), between the
minimum-norm solution and the projector onto the decaying directions. This relation underlies prior numerical methods
for computing this projector, known as the dichotomy projector~\cite{huls2009numerical,froyland2013computing}. The
same shift of the exponents by $s$ underlies the dichotomy (Sacker--Sell) spectrum and the computation of this
spectrum~\cite{sacker1978spectral,huls2010computing}. A version of this relation integrated over $s$ follows from a theorem of
\citet{deninger2011determinants}, which equates the sum of the positive Lyapunov exponents
with a Fuglede--Kadison log-determinant of the propagation operator
(\hyperref[sec:ai]{AI methodology}).

\subsection{Use of AI}
\label{sec:aiuse}

This work was done in collaboration with the AI models GPT-6 Astra and Claude Opus 5.5. GPT-6 Astra produced an
initial derivation. Working with both models, the author then arrived at the more physically transparent derivation
presented in Sec.~\ref{sec:derivation}. Claude Opus 5.5 wrote all code and generated all figures, under the direction of the
author, and assisted in writing and revision of the text. Details are given in \hyperref[sec:ai]{AI methodology}.

\section{Derivation}
\label{sec:derivation}

\subsection{Part 1: Cumulative distribution as a response function}
\label{sec:part1}

We begin by specifying the network and its tangent dynamics. We then show that counting the Lyapunov exponents below
a threshold $s$ amounts to computing a response function of the minimum-norm solution of the shifted tangent dynamics
driven by a source.

We consider the Euler discretization, with time step $0<\delta\leq1$, of the random neural network of \citet{sompolinsky1988chaos},
\begin{align}
    x_i(n+1)&=(1-\delta)\,x_i(n)+\delta\sum_{j=1}^NJ_{ij}\,\phi(x_j(n)),\label{eq:x}\\
    J_{ij}&\overset{\text{i.i.d.}}{\sim}\mathcal N\Big(0,\frac{g^2}{N}\Big),\label{eq:J}
\end{align}
where $g$ is the coupling strength and $\phi(x)=\tanh(x)$ is a nonlinearity. We write $d_i(n)=\phi'(x_i(n))$
for the so-called gains and $D(n)=\operatorname{diag}\big(d_1(n),\ldots,d_N(n)\big)$ for the diagonal matrix of gains
at time $n$. Equation~\eqref{eq:x} has the form $x(n+1)=G(x(n))$ of Sec.~\ref{sec:lyap}, with
$G(x)=(1-\delta)\,x+\delta J\phi(x)$, where $\phi(x)$ acts elementwise on $x$. The Jacobian of $G(x)$ at $x(n)$ is
\begin{equation}
    M(n)=(1-\delta)I_N+\delta JD(n),
\end{equation}
so that a tangent vector evolves as $v(n+1)=M(n)\,v(n)$, and the Lyapunov exponents
$\lambda_1\geq\cdots\geq\lambda_N$ are given by Eq.~\eqref{eq:lyap}. We would like to compute the cumulative
distribution of Lyapunov exponents,
\begin{equation}
    F(s)=\frac1N\,\#\{i:\lambda_i<s\}.
\end{equation}

To count the Lyapunov exponents below $s$, we shift the Lyapunov exponents. In particular, multiplying the Jacobian
$M(n)$ at every step by $e^{-\delta s}$ lowers every Lyapunov exponent by $s$. The shifted tangent dynamics,
\begin{align}
    v(n+1)&=M_s(n)\,v(n),\label{eq:shifted}\\
    M_s(n)&=e^{-\delta s}M(n)\nonumber\\
    &=\alpha_s I_N+\beta_s JD(n),\label{eq:Ms}
\end{align}
with
\begin{align}
    \alpha_s&=(1-\delta)\,e^{-\delta s},\label{eq:alpha}\\
    \beta_s&=\delta\,e^{-\delta s},\label{eq:beta}
\end{align}
therefore have Lyapunov exponents $\lambda_i-s$. Forward in time, the shifted dynamics shrink the directions with
$\lambda_i<s$ exponentially and grow the directions with $\lambda_i>s$ exponentially. Backward in time, the roles are
reversed. Thus, $NF(s)$ is the number of directions that decay forward in time under the shifted dynamics, and the task
is to count these directions.

To do so, we consider tangent trajectories on the infinite window of times $n\in\mathbb Z$. We collect a sequence of
tangent vectors $v(n)$ and a source $I^v(n)$, which drives the step from time $n$ to time $n+1$, into the vectors
\begin{equation}
    \mathcal V=\begin{pmatrix}\vdots\\v(1)\\v(2)\\\vdots\end{pmatrix},\qquad
    \mathcal I^v=\begin{pmatrix}\vdots\\I^v(1)\\I^v(2)\\\vdots\end{pmatrix}.
\end{equation}
The shifted dynamics driven by this source,
\begin{equation}
    v(n+1)=M_s(n)\,v(n)+I^v(n),
    \label{eq:driven}
\end{equation}
where $n\in\mathbb Z$, can then be written as
\begin{equation}
    K\mathcal V=\mathcal I^v,
    \label{eq:KV}
\end{equation}
where $K$ is the block-bidiagonal matrix
\begin{equation}
    K=\begin{pmatrix}
        \ddots & \ddots & & & \\
        & -M_s(1) & I_N & & \\
        & & -M_s(2) & I_N & \\
        & & & \ddots & \ddots
    \end{pmatrix}.
    \label{eq:K}
\end{equation}
For any sequence of tangent vectors $\mathcal V$, the product $K\mathcal V$ collects the residuals of the steps,
\begin{equation}
    (K\mathcal V)(n)=v(n+1)-M_s(n)\,v(n),
    \label{eq:Kaction}
\end{equation}
for $n\in\mathbb Z$. Each block row of $K$ corresponds to one step. We assume that $M(n)$ is invertible for every $n$, so that
Eq.~\eqref{eq:KV} can be solved both forward and backward in time. Since Eq.~\eqref{eq:KV} imposes no initial
condition, the tangent vector $v(n)$ at any single time $n$ can then be chosen freely, and the solutions of
Eq.~\eqref{eq:KV} form an $N$-dimensional affine family, consisting of any particular solution plus a free tangent
trajectory, i.e., a solution of the homogeneous system $K\mathcal V=0$.

Our first step is to remove this freedom, not by fixing the tangent vector at any time, but by selecting the solution
of Eq.~\eqref{eq:KV} with minimum norm, $\lVert\mathcal V\rVert_2^2=\sum_{n\in\mathbb Z}\lVert v(n)\rVert_2^2$. By
linearity, it suffices to consider a source that acts at a single time $k$, i.e., $I^v(k)=w$ and $I^v(n)=0_N$ for
$n\neq k$. A generic solution of Eq.~\eqref{eq:KV} grows exponentially as $n\to\infty$, as $n\to-\infty$, or both. We
now explicitly construct the one solution that decays in both directions, which is the only solution with finite norm
and hence the minimum-norm solution.

We assume that no Lyapunov exponent equals $s$, so that no direction is marginal under the shifted dynamics. The multiplicative ergodic theorem of Oseledets provides, for a typical
trajectory, a splitting of the tangent space at every time $n$ into subspaces, one for each distinct Lyapunov exponent,
such that the norms of the tangent vectors in each subspace change at the rate given by the exponent of that subspace,
both forward and backward in time~\cite{oseledets1968multiplicative}. Grouping these subspaces according to whether
their exponents lie below or above $s$ gives the splitting $\mathbb R^N=E^{\mathrm s}(n)\oplus E^{\mathrm u}(n)$, where
the stable subspace $E^{\mathrm s}(n)$ is spanned by the tangent vectors whose norms change at rates $\lambda_i<s$ and
the unstable subspace $E^{\mathrm u}(n)$ by the tangent vectors whose norms change at rates $\lambda_i>s$. Forward in time, the shifted
dynamics shrink vectors in $E^{\mathrm s}(n)$ exponentially and grow vectors in $E^{\mathrm u}(n)$ exponentially.
Backward in time, the roles are reversed.

The source $w$ enters the tangent vector at time $k+1$, and we decompose $w$ using the subspaces at time $k+1$. Let
$P(k+1)$ be the projector onto $E^{\mathrm s}(k+1)$ along $E^{\mathrm u}(k+1)$, so that $I_N-P(k+1)$ is the
projector onto $E^{\mathrm u}(k+1)$ along $E^{\mathrm s}(k+1)$. We decompose the source into its stable and unstable
components, $P(k+1)\,w$ and $[I_N-P(k+1)]\,w$. Because the splitting is a direct sum, this decomposition exists and is
unique for each $w$. We then construct a solution of Eq.~\eqref{eq:driven}, which we refer to as the constructed solution, that behaves as
follows. Coming in from $n=-\infty$, where the tangent vector vanishes, the tangent vector grows within
$E^{\mathrm u}(n)$ and is mapped to $-[I_N-P(k+1)]\,w$ by the step from time $k$ to time $k+1$. In
the same step, the source $w$ cancels this unstable vector and leaves the stable component $P(k+1)\,w$, which then
decays within $E^{\mathrm s}(n)$ as $n\to\infty$. To compute this solution, we propagate the stable component forward in time from time $k+1$,
and the negative of the unstable component backward in time from time $k+1$,
\begin{equation}
    \begin{aligned}
        &\ \ \vdots\\
        v(k-1)&=-M_s(k-1)^{-1}M_s(k)^{-1}\,[I_N-P(k+1)]\,w,\\
        v(k)&=-M_s(k)^{-1}\,[I_N-P(k+1)]\,w,\\
        v(k+1)&=P(k+1)\,w,\\
        v(k+2)&=M_s(k+1)\,P(k+1)\,w,\\
        &\ \ \vdots
    \end{aligned}
    \label{eq:bounded}
\end{equation}

One can verify that the constructed solution, Eq.~\eqref{eq:bounded}, solves Eq.~\eqref{eq:driven}. Away from the
source, the constructed solution, Eq.~\eqref{eq:bounded}, follows the free shifted dynamics. For example, immediately
before the source,
\begin{align}
    M_s(k-1)\,v(k-1)&=-M_s(k)^{-1}\,[I_N-P(k+1)]\,w\nonumber\\
    &=v(k),
\end{align}
and immediately after the source,
\begin{align}
    M_s(k+1)\,v(k+1)&=M_s(k+1)\,P(k+1)\,w\nonumber\\
    &=v(k+2),
\end{align}
and likewise at all other times away from the source. Across the source,
\begin{align}
    M_s(k)\,v(k)+w&=-[I_N-P(k+1)]\,w+w\nonumber\\
    &=P(k+1)\,w\nonumber\\
    &=v(k+1).
\end{align}
Because the stable component decays forward in time and the negative of the unstable component decays backward in time, the
constructed solution, Eq.~\eqref{eq:bounded}, decays exponentially in both directions and therefore has finite norm.
Every other solution differs from the constructed solution, Eq.~\eqref{eq:bounded}, by a nonzero free tangent
trajectory. Unlike the constructed solution, which switches from the unstable to the stable subspace at the source, a
free tangent trajectory receives no source and thus has a nonzero component that grows exponentially forward or
backward in time. Every other solution therefore has infinite norm, so the constructed solution,
Eq.~\eqref{eq:bounded}, is the only solution of Eq.~\eqref{eq:KV} with finite norm and hence the minimum-norm
solution.

We can now read off the cumulative distribution. The constructed solution, Eq.~\eqref{eq:bounded}, has
$v(k+1)=P(k+1)\,w$, so the one-step response of the minimum-norm solution is the $N\times N$ matrix
$\partial v(k+1)/\partial I^v(k)=P(k+1)$. The causal solution, which
propagates the whole source forward rather than only its stable component, would instead give $I_N$. Since the trace
of a projector is the dimension of the range of the projector, and $\dim E^{\mathrm s}(k+1)=NF(s)$,
\begin{equation}
    F(s)=\frac1N\tr\frac{\partial v(k+1)}{\partial I^v(k)}.
    \label{eq:F}
\end{equation}
Equation~\eqref{eq:F} is exact at finite $N$ for every $s$ that is not a Lyapunov exponent.

\subsection{Part 2: Regularized forward--backward system}
\label{sec:dyn}

To compute the trace in Eq.~\eqref{eq:F} at large $N$, we recast the minimum-norm solution in a form to which the cavity method
applies. We do so in two steps. First, we write the minimum-norm solution as the limit of the unique solution of a
regularized least-squares problem as the regularization vanishes. Second, we write this regularized problem as a
forward--backward dynamical system for two fields, a forward field that runs forward in time and a backward field that
runs backward in time. The restriction of the solution of this system to the forward field solves the regularized problem.

We denote the regularization by $\eta^2$ and call the minimizer of the regularized least-squares cost the regularized
solution,
\begin{equation}
    \mathcal V=\operatorname*{arg\,min}_{\mathcal V'}\Big(\big\lVert K\mathcal V'-\mathcal I^v\big\rVert_2^2+\eta^2\,\lVert\mathcal V'\rVert_2^2\Big).
    \label{eq:Veta}
\end{equation}
The minimizer is unique, since the penalty $\eta^2\lVert\mathcal V'\rVert_2^2$ makes the cost strictly convex. We
write
\begin{equation}
    R^{vv}(n,n')=\frac{\partial v(n)}{\partial I^v(n')}
    \label{eq:Rvvdef}
\end{equation}
for the response of the regularized solution, Eq.~\eqref{eq:Veta}, an $N\times N$ matrix for each pair of times. As
$\eta\to0^+$, the regularized solution, Eq.~\eqref{eq:Veta}, converges to the minimum-norm solution, as for any regularized least-squares
problem that has a solution of finite norm. Thus, Eq.~\eqref{eq:F} becomes
\begin{equation}
    F(s)=\lim_{\eta\to0^+}\frac1N\tr R^{vv}(k+1,k).
    \label{eq:Feta}
\end{equation}
All responses below are evaluated at fixed $\eta>0$ and depend on $\eta$, as they do on $s$.

Setting the gradient of the cost to zero shows that the regularized solution, Eq.~\eqref{eq:Veta}, satisfies
\begin{equation}
    K^T(K\mathcal V-\mathcal I^v)+\eta^2\mathcal V=0.
\end{equation}
To turn this into a dynamical system, we introduce the rescaled residual
\begin{equation}
    \mathcal U=\begin{pmatrix}\vdots\\u(1)\\u(2)\\\vdots\end{pmatrix}=\frac{\mathcal I^v-K\mathcal V}{\eta}.
\end{equation}
In terms of $\mathcal U$, the condition for the regularized solution, Eq.~\eqref{eq:Veta}, becomes the pair of equations
\begin{align}
    \eta\,\mathcal U+K\mathcal V&=\mathcal I^v,\\
    -K^T\mathcal U+\eta\,\mathcal V&=0.
\end{align}
Note that, at $\eta=0$, these two equations decouple, and the forward field $\mathcal V$ then obeys
Eq.~\eqref{eq:KV}, with the freedom of Part~1 unresolved, so that the two equations do not select a solution. For
$\eta>0$, the forward and backward fields are coupled at every time,
and this coupling selects the regularized solution, Eq.~\eqref{eq:Veta}. As $\eta\to0^+$, the backward field vanishes.

We add a second source $\mathcal I^u$ to the right-hand side of the second equation. The readout of the cumulative distribution $F(s)$, Eq.~\eqref{eq:Feta}, does not use the
source $\mathcal I^u$, but
the cavity calculation of Part~3 (Sec.~\ref{sec:part3}) does. In block-matrix form,
\begin{equation}
    B
    \begin{pmatrix}\mathcal U\\\mathcal V\end{pmatrix}
    =\begin{pmatrix}\mathcal I^v\\\mathcal I^u\end{pmatrix},
    \label{eq:B}
\end{equation}
where
\begin{equation}
    B=\begin{pmatrix}\eta I&K\\-K^T&\eta I\end{pmatrix},
    \label{eq:Bdef}
\end{equation}
with $I$ the identity on infinite sequences of $N$-dimensional vectors. The matrix $B$ is $\eta$ times the identity
plus an antisymmetric matrix, so $z^TBz=\eta\lVert z\rVert_2^2$ for every $z$. For $\eta>0$, $B$ is therefore positive
definite and hence invertible, so Eq.~\eqref{eq:B} has a unique solution. At $\mathcal I^u=0$, the restriction of this
solution to the forward field $\mathcal V$ is the regularized solution, Eq.~\eqref{eq:Veta}. This construction is
closely related to Hermitization, a standard tool for the spectra of non-Hermitian random
matrices~\cite{feinberg1997non}.

Written out, the first block row of Eq.~\eqref{eq:B} is a forward equation for $v(n)$, and the second is a backward
equation for $u(n)$. Together with the network, which drives both through the gains, the full system is
\begin{align}
    x_i(n+1)&=(1-\delta)\,x_i(n)+\delta\Big[\sum_{j=1}^NJ_{ij}\,\phi(x_j(n))+I^x_i(n)\Big],\label{eq:x_full}\\
    v_i(n+1)&=\alpha_s\,v_i(n)+\beta_s\sum_{j=1}^NJ_{ij}\,d_j(n)\,v_j(n)-\eta\,u_i(n)+I^v_i(n),\label{eq:v_full}\\
    u_i(n-1)&=\alpha_s\,u_i(n)+\beta_s\,d_i(n)\sum_{j=1}^NJ_{ji}\,u_j(n)+\eta\,v_i(n)-I^u_i(n),\label{eq:u_full}
\end{align}
for all $n\in\mathbb Z$. On the infinite window, this system is statistically stationary. We have also given the
network a source $I^x_i(n)$, which the cavity calculation uses to
define the network's own response; in the physical dynamics, the source $I^x_i(n)$ is zero.

Generalizing Eq.~\eqref{eq:Rvvdef}, we write $R^{ab}(n,n')=\partial a(n)/\partial I^b(n')$, with $a,b\in\{x,u,v\}$, for the
$N\times N$ response matrices of this system, and $R^{ab}_{ij}(n,n')$ for the entries of these matrices. When all three
sources vanish, the forward and backward fields vanish, $v(n)=u(n)=0_N$, while the neurons follow the chaotic network
dynamics. 

\subsection{Part 3: Cavity method and single-site theory}
\label{sec:part3}

We now compute the response $R^{vv}(k+1,k)$ at large $N$ with the cavity method, applied jointly to the regularized forward--backward
system of Part~2 (Sec.~\ref{sec:dyn}) and the network dynamics. To do so, we add a new neuron with index $0$ to a network of $N$ neurons with indices $i \in \{1, \ldots, N\}$,
which we call the reservoir. We compute how neuron $0$ perturbs the reservoir and how the perturbed reservoir acts
back on neuron $0$, then close the resulting equations self-consistently.

For a large network, Eq.~\eqref{eq:Feta} gives the cumulative distribution
\begin{equation}
    \lim_{N\to\infty}F(s)=\lim_{N\to\infty}\lim_{\eta\to0^+}\frac1N\tr R^{vv}(k+1,k).
    \label{eq:FNorder}
\end{equation}
The cavity calculation below is carried out at fixed $\eta>0$, where the forward--backward system of Part~2
(Sec.~\ref{sec:dyn}) has a unique solution. We therefore take $N\to\infty$ before $\eta\to0^+$,
\begin{equation}
    \lim_{N\to\infty}F(s)=\lim_{\eta\to0^+}\lim_{N\to\infty}\frac1N\tr R^{vv}(k+1,k),
    \label{eq:FN}
\end{equation}
assuming that the limits $N\to\infty$ and $\eta\to0^+$ can be exchanged. This exchange is the main assumption of the calculation. It is supported by the
agreement with simulations (Sec.~\ref{sec:results}); the analogous exchange is routinely assumed in random-matrix theory, both for
Hermitian matrices~\cite{potters2021first} and, through Hermitization, for non-Hermitian matrices~\cite{feinberg1997non}. Below, $F(s)$ denotes the large-$N$ limit, Eq.~\eqref{eq:FN}.

As is standard in the cavity method, we assume that the normalized trace in Eq.~\eqref{eq:FN} is self-averaging, so
that at large $N$ it equals the response of a typical neuron. It therefore
suffices to compute the response of neuron $0$ to the sources $I^v_0(n)$ and $I^u_0(n)$ of neuron $0$, with the sources
of the reservoir neurons set to zero.

The reservoir neurons have indices $i,j\in\{1,\ldots,N\}$. Without neuron $0$, the reservoir receives no source, so the
forward and backward fields of the reservoir vanish (Part~2, Sec.~\ref{sec:dyn}). The couplings of neuron $0$, $J_{0i}$ and $J_{i0}$, are drawn
i.i.d.\ from $\mathcal N(0,g^2/N)$, independently of the reservoir couplings $J_{ij}$ and hence of the reservoir
dynamical variables. The self-coupling $J_{00}$ gives terms of order $N^{-1/2}$, which we drop. Below, $x_i(n)$, $d_i(n)$, and the response functions $R^{ab}_{ij}(n,n')$, including the
network's own response $R^{xx}_{ij}(n,n')$, refer to the reservoir \emph{without} neuron $0$.

The introduction of neuron $0$ perturbs the reservoir. In particular, neuron $0$ adds the input
$\delta J_{i0}\phi(x_0(n))$ to Eq.~\eqref{eq:x_full}, the input $\beta_s J_{i0}d_0(n)v_0(n)$ to Eq.~\eqref{eq:v_full}, and
the input $\beta_s d_i(n)J_{0i}u_0(n)$ to Eq.~\eqref{eq:u_full}. Each of these additional inputs is
$\order{N^{-1/2}}$ and enters the reservoir equations like a source, so at leading order in $N^{-1/2}$ the
reservoir responds linearly, through the reservoir responses $R^{ab}_{ij}(n,n')$. The states of the reservoir
neurons are perturbed by $\Delta x_i(n)$, while the forward and backward fields of the reservoir, which were zero, acquire values
$v_i(n)$ and $u_i(n)$ that are $\order{N^{-1/2}}$,
\begin{align}
    \Delta x_i(n)&=\sum_{j=1}^N\sum_{n'\in\mathbb Z}R^{xx}_{ij}(n,n')\,J_{j0}\,\phi(x_0(n')),\label{eq:dx}\\
    v_i(n)&=\beta_s\sum_{j=1}^N\sum_{n'\in\mathbb Z}\left[R^{vv}_{ij}(n,n')\,J_{j0}\,d_0(n')\,v_0(n')
        -R^{vu}_{ij}(n,n')\,d_j(n')\,J_{0j}\,u_0(n')\right],\label{eq:dv}\\
    u_i(n)&=\beta_s\sum_{j=1}^N\sum_{n'\in\mathbb Z}\left[R^{uv}_{ij}(n,n')\,J_{j0}\,d_0(n')\,v_0(n')
        -R^{uu}_{ij}(n,n')\,d_j(n')\,J_{0j}\,u_0(n')\right].\label{eq:du}
\end{align}
By contrast, the fields $v_0(n)$ and $u_0(n)$ of neuron $0$, which the sources $I^v_0(n)$ and $I^u_0(n)$ drive directly
(Eqs.~\eqref{eq:v0} and~\eqref{eq:u0} below), are $\order{1}$.

The perturbations $\Delta x_i(n)$ also perturb the gains, by $\Delta d_i(n)=\phi''(x_i(n))\,\Delta x_i(n)$, which is
$\order{N^{-1/2}}$. However, these gain perturbations contribute to the equations for $v_0(n)$ and $u_0(n)$,
Eqs.~\eqref{eq:v0} and~\eqref{eq:u0} below, only as sums of $N$ terms, each $\order{N^{-3/2}}$. These sums are at most
$\order{N^{-1/2}}$, even if the terms add coherently, and are thus negligible relative to the $\order{1}$ terms of these
equations. We therefore drop the gain perturbations.

These perturbations act back on neuron $0$. Inserting Eqs.~\eqref{eq:dx}--\eqref{eq:du} into Eqs.~\eqref{eq:x_full}--\eqref{eq:u_full} for neuron $0$, with $\phi(x_i(n))$
expanded to first order in $\Delta x_i(n)$, gives
\begin{align}
    x_0(n+1)&=(1-\delta)\,x_0(n)+\delta\Big[\xi_0(n)+\sum_{n'\in\mathbb Z}\Gamma^{xx}(n,n')\,\phi(x_0(n'))\Big],\label{eq:x0}\\
    v_0(n+1)&=\alpha_s\,v_0(n)-\eta\,u_0(n)+\sum_{n'\in\mathbb Z}\Big[\Gamma^{vv}(n,n')\,d_0(n')\,v_0(n')-\Gamma^{vu}(n,n')\,u_0(n')\Big]+I^v_0(n),\label{eq:v0}\\
    u_0(n-1)&=\alpha_s\,u_0(n)+\eta\,v_0(n)+d_0(n)\sum_{n'\in\mathbb Z}\Big[\Gamma^{uv}(n,n')\,d_0(n')\,v_0(n')-\Gamma^{uu}(n,n')\,u_0(n')\Big]-I^u_0(n),\label{eq:u0}
\end{align}
where the neuronal cavity field is
\begin{equation}
    \xi_0(n)=\sum_{i=1}^NJ_{0i}\,\phi(x_i(n)),
    \label{eq:cavfield}
\end{equation}
and the reaction kernels are
\begin{align}
    \Gamma^{xx}(n,n')&=\sum_{i,j=1}^NJ_{0i}\,d_i(n)\,R^{xx}_{ij}(n,n')\,J_{j0},\label{eq:Gxx}\\
    \Gamma^{vv}(n,n')&=\beta_s^2\sum_{i,j=1}^NJ_{0i}\,d_i(n)\,R^{vv}_{ij}(n,n')\,J_{j0},\\
    \Gamma^{vu}(n,n')&=\beta_s^2\sum_{i,j=1}^NJ_{0i}\,d_i(n)\,R^{vu}_{ij}(n,n')\,d_j(n')\,J_{0j},\\
    \Gamma^{uv}(n,n')&=\beta_s^2\sum_{i,j=1}^NJ_{i0}\,R^{uv}_{ij}(n,n')\,J_{j0},\\
    \Gamma^{uu}(n,n')&=\beta_s^2\sum_{i,j=1}^NJ_{i0}\,R^{uu}_{ij}(n,n')\,d_j(n')\,J_{0j}.\label{eq:Guu}
\end{align}

We now take $N$ to be large. By the cavity construction, the couplings of neuron $0$ are independent of the reservoir
quantities that these couplings multiply in Eqs.~\eqref{eq:cavfield}--\eqref{eq:Guu}. Together with the number of row couplings $J_{0i}$ and column couplings $J_{i0}$
that each quantity contains, this independence determines the large-$N$ behavior of each quantity.
\begin{itemize}
\item The neuronal cavity field $\xi_0(n)$ is linear in the row couplings $J_{0i}$. Since these couplings are
independent of the reservoir activations $\phi(x_i(n))$, the neuronal cavity field is Gaussian, with mean zero and
covariance $(g^2/N)\sum_{i=1}^N\phi(x_i(n))\,\phi(x_i(n'))$.
\item The kernels $\Gamma^{xx}(n,n')$, $\Gamma^{vv}(n,n')$, and $\Gamma^{uu}(n,n')$ contain one row and one column
coupling, $J_{0i}J_{j0}$. Since the row and column couplings are independent of each other and of the reservoir
quantities, these kernels have mean zero and are $\order{N^{-1/2}}$, so we drop them.
\item The kernels $\Gamma^{vu}(n,n')$ and $\Gamma^{uv}(n,n')$ contain two row couplings, $J_{0i}J_{0j}$, and two
column couplings, $J_{i0}J_{j0}$, respectively. Since these couplings are independent of the reservoir quantities,
\begin{equation}
    \sum_{i,j=1}^NJ_{0i}J_{0j}\,Z_{ij}\to\frac{g^2}{N}\sum_{i=1}^NZ_{ii}
\end{equation}
for reservoir quantities $Z_{ij}$, and likewise for column couplings. These kernels therefore concentrate to $\order{1}$ quantities,
\begin{align}
    \Gamma^{vu}(n,n')&\to A(n,n'),\\
    \Gamma^{uv}(n,n')&\to\gamma_s\,P(n,n'),\label{eq:GP}
\end{align}
where $\gamma_s=\beta_s^2g^2$ and
\begin{align}
    A(n,n')&=\frac{\gamma_s}{N}\sum_{i=1}^Nd_i(n)\,R^{vu}_{ii}(n,n')\,d_i(n'),\label{eq:A}\\
    P(n,n')&=\frac1N\sum_{i=1}^NR^{uv}_{ii}(n,n').\label{eq:P}
\end{align}
\end{itemize}
Neuron $0$ therefore obeys
\begin{align}
    x_0(n+1)&=(1-\delta)\,x_0(n)+\delta\,\xi_0(n),\label{eq:x0lim}\\
    v_0(n+1)&=\alpha_s\,v_0(n)-\eta\,u_0(n)-\sum_{n'\in\mathbb Z}A(n,n')\,u_0(n')+I^v_0(n),\\
    u_0(n-1)&=\alpha_s\,u_0(n)+\eta\,v_0(n)+\gamma_s\,d_0(n)\sum_{n'\in\mathbb Z}P(n,n')\,d_0(n')\,v_0(n')-I^u_0(n).
\end{align}
Thus, the effect of the reservoir on neuron $0$ is captured by the statistics of the neuronal cavity field
$\xi_0(n)$ and by the two kernels $A(n,n')$ and $P(n,n')$. It remains to compute these quantities, which we now do
self-consistently.

Note that neuron $0$ is statistically equivalent to any reservoir neuron. The normalized sums over the reservoir, in
the covariance of $\xi_0(n)$ and in Eqs.~\eqref{eq:A}--\eqref{eq:P}, are therefore averages over the single-site
Gaussian process just derived. We write $\tavg{\cdot}_{x_0}$ for the average over this process.

The single-site problem has a one-way structure in which the neuronal dynamics of neuron $0$,
Eq.~\eqref{eq:x0lim}, do not involve the forward and backward fields and are those of the original dynamical
mean-field theory~\cite{sompolinsky1988chaos,crisanti2018path}.
This mirrors the high-dimensional network, whose trajectory drives the
tangent dynamics but not vice versa. Specifically, the neuronal cavity field $\xi_0(n)$ is
a zero-mean Gaussian process whose covariance is determined by the self-consistency condition
\begin{equation}
    \tavg{\xi_0(n)\,\xi_0(n')}_{x_0}=g^2\tavg{\phi(x_0(n))\,\phi(x_0(n'))}_{x_0}.
    \label{eq:dmft}
\end{equation}
The solution of Eq.~\eqref{eq:dmft} determines the statistics of the gain trajectories
$d_0(n)=\phi'(x_0(n))$, which then drive the forward and backward fields of neuron $0$.

For the forward and backward fields of neuron $0$, we write the equations as a $2\times2$ block system, as in
Eq.~\eqref{eq:B}, but with time indices only. We collect the single-site sequences into the vectors
\begin{equation}
    \mathcal V_0=\begin{pmatrix}\vdots\\v_0(1)\\v_0(2)\\\vdots\end{pmatrix},\qquad
    \mathcal U_0=\begin{pmatrix}\vdots\\u_0(1)\\u_0(2)\\\vdots\end{pmatrix},
\end{equation}
and likewise for the sources $\mathcal I^v_0$ and $\mathcal I^u_0$. We define the bidiagonal matrix
\begin{equation}
    L=\begin{pmatrix}
        \ddots & \ddots & & & \\
        & -\alpha_s & 1 & & \\
        & & -\alpha_s & 1 & \\
        & & & \ddots & \ddots
    \end{pmatrix},
    \label{eq:L}
\end{equation}
which acts as
\begin{align}
    (L\mathcal V_0)(n)&=v_0(n+1)-\alpha_s\,v_0(n),\\
    (L^T\mathcal U_0)(n)&=u_0(n-1)-\alpha_s\,u_0(n).
\end{align}
We also let $D_0$ denote the diagonal matrix with diagonal entries $d_0(n)$, $A$ and $P$ the matrices with entries $A(n,n')$
and $P(n,n')$, and $I$ the identity on single-site sequences. The forward and backward equations of neuron $0$ then
read
\begin{equation}
    \begin{pmatrix}\eta I+A&L\\-L^T&\eta I+\gamma_s\,D_0PD_0\end{pmatrix}
    \begin{pmatrix}\mathcal U_0\\\mathcal V_0\end{pmatrix}
    =\begin{pmatrix}\mathcal I^v_0\\\mathcal I^u_0\end{pmatrix}.
    \label{eq:Bsite}
\end{equation}

We write $R^{ab}_{00}(n,n')$ for the responses of Eq.~\eqref{eq:Bsite}, which depend on the gain trajectory $d_0(n)$.
Inverting Eq.~\eqref{eq:Bsite} by blocks gives these responses in matrix form,
\begin{align}
    R^{vu}_{00}&=\left[\eta I+\gamma_s\,D_0PD_0+L^T(\eta I+A)^{-1}L\right]^{-1},\label{eq:Rvu00}\\
    R^{uv}_{00}&=\left[\eta I+A+L\,(\eta I+\gamma_s\,D_0PD_0)^{-1}L^T\right]^{-1},\label{eq:Ruv00}\\
    R^{vv}_{00}&=R^{vu}_{00}\,L^T(\eta I+A)^{-1}.\label{eq:Rvv00}
\end{align}

The kernels are then determined self-consistently by
\begin{subequations}\label{eq:closed}
\begin{align}
    A(n,n')&=\gamma_s\tavg{d_0(n)\,R^{vu}_{00}(n,n')\,d_0(n')}_{x_0},\\
    P(n,n')&=\tavg{R^{uv}_{00}(n,n')}_{x_0},
\end{align}
\end{subequations}
where $d_0(n)=\phi'(x_0(n))$. Note that the gains $d_0(n)$ and $d_0(n')$ appear inside the average defining $A(n,n')$, but outside the average defining $P(n,n')$, where they instead multiply $P(n,n')$ in Eq.~\eqref{eq:Bsite}. As $\eta$ decreases, we follow the solution of Eq.~\eqref{eq:closed} continuously from large $\eta$, where the kernels
are small. The cumulative
distribution is then read out from Eq.~\eqref{eq:FN} as
\begin{equation}
    F(s)=\lim_{\eta\to0^+}\tavg{R^{vv}_{00}(k+1,k)}_{x_0}.
    \label{eq:Fsite}
\end{equation}
The parameter $s$ enters only through $\alpha_s$ and $\gamma_s$. In the stationary state, the kernels $A(n,n')$ and
$P(n,n')$, the gain correlations, and the averaged responses depend only on $n-n'$; in particular,
Eq.~\eqref{eq:Fsite} does not depend on $k$. The responses $R^{ab}_{00}(n,n')$ for a single gain trajectory, in contrast, depend on $n$ and $n'$ separately, because $D_0$ is not translation invariant. The kernels are also symmetric, $A(n,n')=A(n',n)$ and
$P(n,n')=P(n',n)$, since solving Eq.~\eqref{eq:B} by blocks gives the symmetric matrices
$R^{vu}=(\eta I+K^TK/\eta)^{-1}$ and $R^{uv}=(\eta I+KK^T/\eta)^{-1}$.

\section{Results}
\label{sec:results}

\subsection{Trivial fixed point and maximum Lyapunov exponent}
\label{sec:limits}

We first check the single-site theory against two known results, the Lyapunov spectrum at the trivial fixed point and
the maximum Lyapunov exponent in the chaotic phase. To our knowledge, these results have appeared in the literature only as
$\delta\to0$ and at $\delta=1$, but both extend directly to finite $\delta$ (Table~\ref{tab:limits}). In both cases, the
single-site theory simplifies and recovers the expected results. Details are given in Appendices~\ref{app:tfp} and~\ref{app:maxexp}.

\begin{table}[h]
    \centering
    \caption{Trivial fixed point and maximum Lyapunov exponent. The first row gives the region of eigenvalues $\mu$ of
    $J$ in which $\lambda_i<s$ at the trivial fixed point, so that $F(s)$ is the fraction of the disk $|\mu|<g$ that
    lies in this region. The second row gives the maximum Lyapunov exponent $\lambda_1$ in the chaotic phase.}
    \label{tab:limits}
    \begin{tabular}{lccc}
        \hline
        & finite $\delta$ & $\delta\to0$ & $\delta=1$\\
        \hline
        Trivial fixed point, region with $\lambda_i<s$ & $|\mu-c_\delta|<r_s$ & $\re\mu<1+s$ & $|\mu|<e^s$\\
        Chaotic phase, $\lambda_1$ & $s_*$ & $-1+\sqrt{1-E_0}$ & $\log\big(g\sqrt{C^d(0)}\big)$\\
        \hline
    \end{tabular}
\end{table}

At the trivial fixed point, $x_i(n)=0$, which is stable for $g<1$, every gain is $d_i(n)=1$, and the network is
linear, with the constant Jacobian $M(n)=(1-\delta)I_N+\delta J$. The Lyapunov exponents are then
$\lambda_i=\delta^{-1}\log|1-\delta+\delta\mu_i|$, where $\mu_i$ are the eigenvalues of $J$. As $N\to\infty$, these
eigenvalues are uniform on the disk $|\mu|<g$, by the circular
law~\cite{ginibre1965statistical,girko1985circular,sommers1988spectrum}. Thus, $F(s)$ is the fraction of the disk $|\mu|<g$
that lies inside a second disk, $|\mu-c_\delta|<r_s$, in which $\lambda_i<s$, with $c_\delta=1-\delta^{-1}$ and
$r_s=\delta^{-1}e^{\delta s}$ (Fig.~\ref{fig:tfp}a and Appendix~\ref{app:tfp-exact}). As $\delta\to0$, the second disk becomes the half-plane $\re\mu<1+s$, and the
density of Lyapunov exponents $F'(s)$ reduces to the semicircle law noted by \citet{engelken2023lyapunov}
(Fig.~\ref{fig:tfp}b). At $\delta=1$, the two disks are concentric, and $F(s)=e^{2s}/g^2$ for $s<\log g$, as noted by
\citet{curato2013onset}.

As expected, the single-site theory recovers this result. At the trivial fixed point, every gain is constant, and the gains are the
only source of time dependence in the single-site problem. The single-site problem is therefore invariant under
translations in time and can be solved in closed form by a Fourier transform in time. In an interpretation of this
solution in two-dimensional electrostatics, $F(s)$ is the fraction of a uniform charge on the disk $|\mu|<g$ enclosed by
the circle $|\mu-c_\delta|=r_s$, by Gauss's law. Thus, the single-site theory recovers the circular-law result for
every $\delta$ (Fig.~\ref{fig:tfp}c and Appendices~\ref{app:tfp-site} and~\ref{app:tfp-gauss}).

The maximum Lyapunov exponent $\lambda_1$, meanwhile, is known in two limits, in terms of the autocorrelation
$C^d(\tau)=\tavg{d_0(n)\,d_0(n')}_{x_0}$ of the gains at lag $\tau=(n-n')\delta$. In continuous time, $\delta\to0$,
\citet{sompolinsky1988chaos} obtained $\lambda_1=-1+\sqrt{1-E_0}$, where $E_0$ is the ground-state energy of a
Schr\"odinger operator with potential $1-g^2C^d(\tau)$. At $\delta=1$, where the network has no leak,
\citet{molgedey1992suppressing} obtained $\lambda_1=\log\big(g\sqrt{C^d(0)}\big)$. Both results are limits of a single
calculation at finite $\delta$, in which dynamical mean-field theory is applied to two copies of the network
(Appendix~\ref{app:replica}). At finite $\delta$, $\lambda_1=s_*$, where $s_*$ is the largest value of $s$ at which $0$
lies in the spectrum of a linear operator $\mathcal T_s$, Eq.~\eqref{app:T}, which acts on functions of the lag $\tau$
and depends on $s$ through $\alpha_s$ and $\gamma_s$. Since $\mathcal T_s$ depends on $s$, finding $s_*$ is a
nonlinear eigenvalue problem, which becomes linear as $\delta\to0$, where $s$ enters only through the energy of the
Schr\"odinger operator. As $\delta\to0$ and at $\delta=1$, $s_*$ reduces to the results
of \citet{sompolinsky1988chaos} and \citet{molgedey1992suppressing}, respectively (Table~\ref{tab:limits}, Fig.~\ref{fig:maxexp}b,c, and
Appendix~\ref{app:limits}).

Again, as expected, the single-site theory recovers this operator. Above the upper edge of the Lyapunov spectrum,
where $F(s)=1$, the kernels of the single-site theory vanish as $\eta\to0^+$, and the solution with vanishing kernels ceases to exist
exactly where $0$ enters the spectrum of $\mathcal T_s$ (Appendix~\ref{app:edge}).

\subsection{Numerical solution and simulations}
\label{sec:numerics}

To solve the single-site theory numerically, we iterate Eq.~\eqref{eq:closed} on a periodic window of $m$ steps,
averaging over gain trajectories sampled from the self-consistent Gaussian process for $x_0(n)$, Eq.~\eqref{eq:dmft},
and we lower $\eta$ toward zero in a few steps. For $\delta\to0$, we solve the continuous-time limit of the single-site
theory on a time grid. To validate the theory, we compare the numerical solution of the single-site theory,
Eqs.~\eqref{eq:closed} and~\eqref{eq:Fsite}, with simulations of the network, Eq.~\eqref{eq:x}. In the
simulations, we compute the Lyapunov spectrum with the QR method, which propagates a set of tangent vectors
with the Jacobians $M(n)$ and repeatedly reorthonormalizes them~\cite{benettin1980lyapunov,engelken2023lyapunov}. For $\delta\to0$, we integrate the
continuous-time network and its tangent dynamics with the fourth-order Runge--Kutta method. Details of both computations are given in Appendix~\ref{app:numerics}.

\subsection{Lyapunov spectra}
\label{sec:spectra}

\begin{figure}[t]
    \centering
    \includegraphics[width=\linewidth]{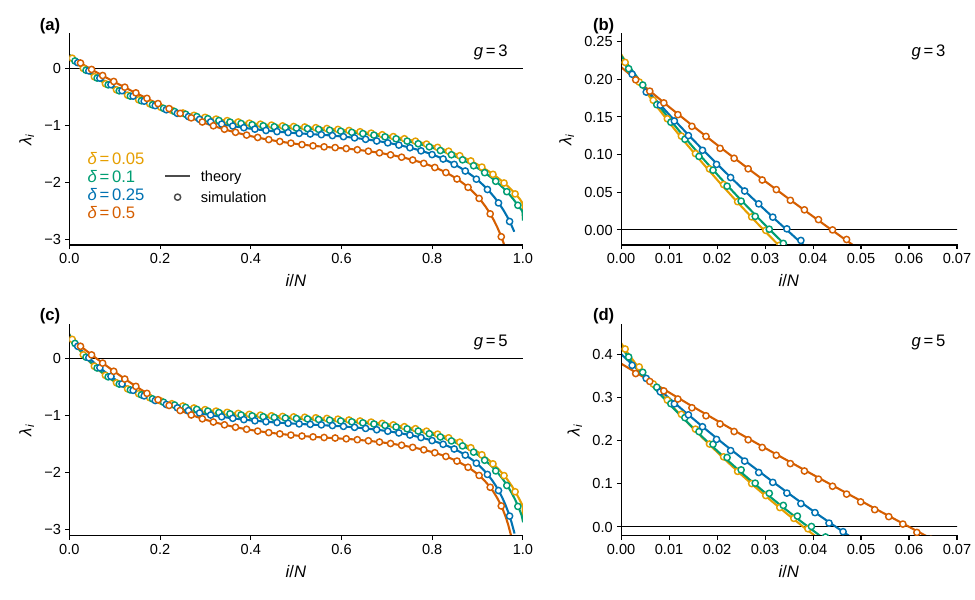}
    \caption{Lyapunov spectra for $\delta=0.05$, $0.1$, $0.25$, and $0.5$, with the legend in (a) for all panels. The exponents
    $\lambda_i$ are plotted against the rank fraction $i/N$. Lines show the single-site theory, Eq.~\eqref{eq:Fsite}, as
    $s$ against $1-F(s)$, ending at the upper edge $s_*$ (Appendix~\ref{app:edge}), and dots show simulations of one
    network with $N=4096$.
    \textbf{(a)} Full Lyapunov spectra for $g=3$, with dots for every 100th Lyapunov exponent.
    \textbf{(b)} Positive Lyapunov exponents of (a), enlarged, with dots for every $\sim$10th Lyapunov exponent.
    \textbf{(c)} Same as (a) for $g=5$.
    \textbf{(d)} Same as (b) for $g=5$.}
    \label{fig:spectra}
\end{figure}

Figure~\ref{fig:spectra} shows the Lyapunov spectra for two values of $g$ and several time steps $\delta$. In the theory, the
exponent at rank fraction $i/N$ is the value of $s$ at which $1-F(s)=i/N$, with $F(s)$ given by
Eq.~\eqref{eq:Fsite}. The theory and the simulations agree closely across the computed range of $s$.

\subsection{Attractor dimension, entropy rate, and participation-ratio dimensions}
\label{sec:dimension}

\begin{figure}[t]
    \centering
    \includegraphics[width=\linewidth]{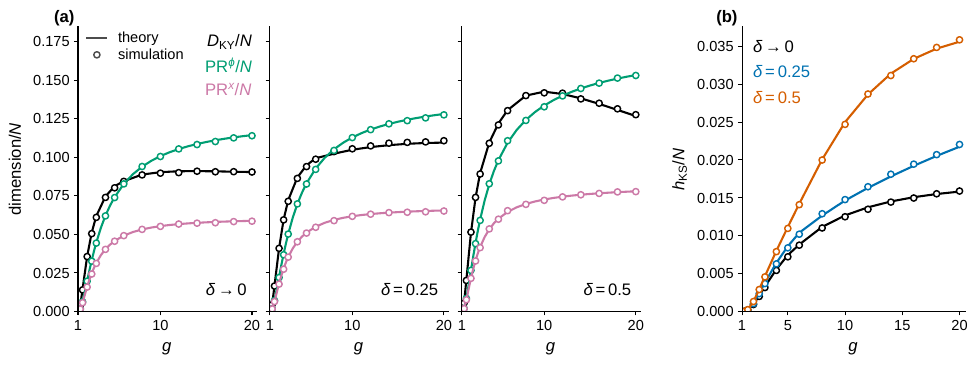}
    \caption{Attractor dimension, participation-ratio dimensions, and entropy rate as functions of the coupling strength
    $g$, with the line and marker legend in (a) for both panels. Lines show theory, and dots show simulations with
    $N=4096$ (the median over five larger networks near the onset of chaos) for $D_{\mathrm{KY}}/N$ and
    $h_{\mathrm{KS}}/N$ and the mean over two networks with $N=16384$ for $\mathrm{PR}^\phi/N$ and $\mathrm{PR}^x/N$
    (Appendix~\ref{app:num-sim}).
    \textbf{(a)} Kaplan--Yorke dimension $D_{\mathrm{KY}}/N$, Eq.~\eqref{eq:dky_theory}, and participation ratios
    $\mathrm{PR}^\phi/N$ and $\mathrm{PR}^x/N$, Eq.~\eqref{eq:PRdef}, for $\delta\to0$, $0.25$, and $0.5$ (left to
    right). The theory is the single-site theory for $D_{\mathrm{KY}}/N$ and the two-site cavity theory
    (Appendix~\ref{app:pr}) for $\mathrm{PR}^\phi/N$ and $\mathrm{PR}^x/N$.
    \textbf{(b)} Entropy rate $h_{\mathrm{KS}}/N$, Eq.~\eqref{eq:hks_theory}, for $\delta\to0$, $0.25$, and $0.5$, with
    the legend in the panel. The theory is the single-site theory.}
    \label{fig:dimension_entropy}
\end{figure}

At large $N$, sums over exponents become integrals over the density of exponents $F'(s)$. Thus, the Kaplan--Yorke
dimension, Eq.~\eqref{eq:ky}, and the entropy rate, given by the sum of the positive exponents
(Sec.~\ref{sec:dimension_entropy}), are
\begin{align}
    \frac{D_{\mathrm{KY}}}{N}&=1-F(s_{\mathrm{KY}}),\label{eq:dky_theory}\\
    \frac{h_{\mathrm{KS}}}{N}&=\int_0^\infty s\,F'(s)\,ds,\label{eq:hks_theory}
\end{align}
where $s_{\mathrm{KY}}<0$ solves $\int_{s_{\mathrm{KY}}}^\infty s\,F'(s)\,ds=0$.

For comparison, we also compute the participation-ratio dimensions of the preactivations and the firing rates, which were the focus of earlier work~\cite{clark2023dimension,clark2025connectivity}. For $a\in\{x,\phi\}$, with
$\phi_i(n)=\phi(x_i(n))$, let
\begin{equation}
    C^a_{ij}(\tau)=\tavg{a_i(n)\,a_j(n')}_n,\qquad \tau=(n-n')\delta,
\end{equation}
be the covariance at lag $\tau$, where $\tavg{\cdot}_n$ denotes the average over $n$ at fixed $n-n'$ (the time averages
of $a_i(n)$ vanish by the symmetry $x\to-x$ of the network). The
participation ratio is
\begin{equation}
    \mathrm{PR}^a=\frac{\Big(\sum_{i=1}^NC^a_{ii}(0)\Big)^2}{\sum_{i,j=1}^NC^a_{ij}(0)^2},
    \label{eq:PRdef}
\end{equation}
which is the participation-ratio dimension discussed in Sec.~\ref{sec:related}. At large $N$, the diagonal covariances
$C^a_{ii}(\tau)$ approach the single-site autocovariance $C^a(\tau)=\tavg{a_0(n)\,a_0(n')}_{x_0}$, with
$\phi_0(n)=\phi(x_0(n))$. The off-diagonal covariances $C^a_{ij}(\tau)$, $i\neq j$, are
$\order{N^{-1/2}}$, but the $N(N-1)$ off-diagonal covariances together contribute at the same order as the $N$
diagonal covariances, through the four-point function
\begin{equation}
    \psi^a(\tau_1,\tau_2)=\frac1N\sum_{\substack{i,j=1\\i\neq j}}^NC^a_{ij}(\tau_1)\,C^a_{ij}(\tau_2),
\end{equation}
which concentrates to a value of order one as $N\to\infty$. Thus, like $D_{\mathrm{KY}}/N$ and $h_{\mathrm{KS}}/N$,
$\mathrm{PR}^a/N$ converges to a self-averaging value of order one,
$\mathrm{PR}^a/N\to C^a(0)^2/\big[C^a(0)^2+\psi^a(0,0)\big]$.

The four-point function $\psi^a(\tau_1,\tau_2)$ follows from the two-site cavity derivation of \citet{clark2023dimension}, in which two neurons are added to the reservoir, or equivalently from
path-integral fluctuations~\cite{clark2025connectivity}. The two-site cavity derivation carries over straightforwardly to
finite time step $\delta$ (Appendix~\ref{app:pr}).

Figure~\ref{fig:dimension_entropy} shows the attractor dimension and the participation ratios
(Fig.~\ref{fig:dimension_entropy}a) and the entropy rate (Fig.~\ref{fig:dimension_entropy}b) as functions of $g$ for
several time steps $\delta$. The theory and the simulations again agree closely. At every time step, the two
participation ratios and the entropy rate increase monotonically with $g$ over the range of $g$ shown, and the
participation ratios grow more slowly at strong coupling, as also found numerically by
\citet{engelken2023lyapunov}. The
attractor dimension saturates at strong coupling and, as found numerically~\cite{engelken2023lyapunov}, eventually
decreases. Over the range of $g$ shown, this decrease is visible at $\delta=0.5$, where the attractor dimension has a
maximum near $g=10$.

\FloatBarrier

\section{Discussion}
\label{sec:discussion}

We derived the full Lyapunov spectrum of a random neural network at large $N$. The derivation combines an exact
identity at finite $N$, which expresses the cumulative distribution of Lyapunov exponents as the one-step response of
the minimum-norm solution of the shifted tangent dynamics, with a cavity calculation that reduces this response to a
self-consistent single-site problem. The resulting theory agrees closely with simulations. Together with recent
calculations of the participation-ratio dimension~\cite{clark2023dimension,clark2025connectivity}, all of the
quantities in Fig.~\ref{fig:dimension_entropy} can now be computed from theory.

The main assumption of the derivation is that the limit of large $N$ and the limit of vanishing regularization can be
exchanged. A more mathematical treatment would need to justify this exchange, along with other steps of the cavity
calculation, a task beyond the expertise of the author. Such a treatment could also enable a formalization of the
result in a proof assistant.

The approach extends straightforwardly to rate networks with structure beyond i.i.d.\ couplings, such as networks
with several populations of neurons, for example, separate excitatory and inhibitory populations; networks whose
couplings have a low-rank component; and networks with correlations between reciprocal couplings $J_{ij}$ and $J_{ji}$.
One could also attempt an extension to networks of spiking neurons. Beyond recurrent
networks, it would be intriguing to compute the Lyapunov spectra of the reasoning dynamics of looped
transformers~\cite{lai2026fractal} and of the dynamics of the weights of neural networks during
learning~\cite{ostrow2023beyond}.

\section*{Acknowledgments}
\addcontentsline{toc}{section}{Acknowledgments}
\label{sec:ack}

The author thanks Ashok Litwin-Kumar for proposing this problem as a target for AI-assisted solution and for
insightful exchanges; Jacob Zavatone-Veth, Alexander van Meegen, and Jonathan Kadmon for enjoyable discussions at
the 2026 Bernstein Conference; and Rainer Engelken for first introducing the author to Lyapunov spectra and their
numerical computation. The author is supported by the Flatiron Institute, a division of the Simons Foundation.

\section*{AI methodology}
\addcontentsline{toc}{section}{AI methodology}
\label{sec:ai}

\paragraph{Initial derivation}
The author had previously assumed that the present problem was analytically intractable.
The initial derivation of a self-consistent single-site problem whose solution matched large-$N$ simulations was
produced by GPT-6 Astra with Max thinking. The prompt below was given within an existing session in which the model
had previously derived other results on the theory of random recurrent neural networks. The model then worked
independently for 100 minutes following the prompt (typos corrected),
\begin{quote}
``Great work so far. Next task: Derive the full $N\to\infty$ Lyapunov spectrum of the SCS model (see Engelken for
numerics). Use whichever technique gets the result and is the most elegant/transparent. Ensure that your writeup is
intelligible. Do not stop until you have nailed the answer. You must complete this task.''
\end{quote}
That derivation was carried out for the continuous-time network. It shifted the Lyapunov exponents by $s$, as in
this paper, and introduced the potential $p_N(s)=N^{-1}\sum_{i=1}^N\max(\lambda_i-s,0)$, whose derivative with
respect to $s$ is $F(s)-1$. The central step was a theorem of \citet{deninger2011determinants} (Theorem~9). For a stationary
ergodic sequence of invertible matrices, such as the Jacobians along a trajectory, this theorem equates the sum of the
positive Lyapunov exponents with the Fuglede--Kadison log-determinant of the operator that propagates tangent vectors
along the trajectory. By taking the time step to zero in this theorem, the original derivation expressed $p_N(s)$, up to a term linear in $s$ and a
constant, as the Fuglede--Kadison log-determinant of the first-order time operator $\partial_t+1+s-JD(t)$, which acts
on neuron and time indices. This log-determinant is defined with a positive regularizer that is taken to zero at the
end. Differentiating with respect to $s$ at fixed regularizer gave a regularized inverse of this time operator. The original derivation computed this regularized inverse
by Hermitizing the time operator into a $2\times2$ block system and averaging over the Gaussian couplings with a matrix
Dyson equation. It handled the dependence of the gains on the couplings by a Gaussian conditioning argument, which decouples the gains from the
couplings in the product of Jacobians. In particular, conditioning on the network trajectory over a fixed time window replaces the
couplings by independent Gaussian couplings plus a correction of rank at most the number of time steps, which does not
affect $F(s)$ as $N\to\infty$. As in the present paper, the exchange of the large-$N$ limit with the
long-time and zero-regularizer limits was assumed.

To relate the original derivation to the present paper, we restate its central step in the discrete-time notation of
Part~1 (Sec.~\ref{sec:part1}). Applied to the shifted Jacobians $M_s(n)$, Deninger's theorem reads
\begin{equation}
    p_N(s)=\frac{1}{N\delta}\log{\det}_{\mathrm{FK}}K,
    \label{eq:deninger}
\end{equation}
where $K$ is the matrix of Eq.~\eqref{eq:K}. The Fuglede--Kadison log-determinant is
\begin{equation}
    \log{\det}_{\mathrm{FK}}K=\lim_{\eta\to0^+}\lim_{m\to\infty}\frac{1}{2m}\log\det\big(K_m^TK_m+\eta^2I_{mN}\big),
    \label{eq:FKdet}
\end{equation}
where $K_m$ is the part of $K$ that acts on the tangent vectors $v(2),\ldots,v(m+1)$,
\begin{equation}
    K_m=\begin{pmatrix}
        I_N & & & \\
        -M_s(2) & I_N & & \\
        & \ddots & \ddots & \\
        & & -M_s(m) & I_N
    \end{pmatrix}.
\end{equation}
In this expression, the order of the limits is important. In particular, since $K_m$ is block lower triangular with $I_N$ on its diagonal, $\det K_m=1$
for any Jacobians, so that taking $\eta\to0^+$ first would give zero and lose all Lyapunov information. At fixed
regularizer $\eta$, differentiating the regularized log-determinant in Eq.~\eqref{eq:FKdet}
with respect to $s$ produces the regularized inverse
$(K^TK+\eta^2I)^{-1}K^T$, which is the response $R^{vv}(n,n')$, Eq.~\eqref{eq:Rvvdef}, of the regularized solution,
Eq.~\eqref{eq:Veta}. As $\eta\to0^+$, the derivative of the right-hand side of Eq.~\eqref{eq:deninger} with
respect to $s$ thus reduces to $F(s)-1$, through the one-step response of Eq.~\eqref{eq:F}, and therefore equals the
derivative of $p_N(s)$. Since both sides of Eq.~\eqref{eq:deninger} vanish as $s\to\infty$, the two sides agree for
every $s$.

\paragraph{Route to the present paper}
The author found this derivation intriguing, albeit complicated and somewhat mysterious, since Deninger's theorem,
particularly in its continuous-time form, gave no direct picture of why the $2\times2$ block system, as the
regularizer vanishes, counts the Lyapunov exponents below $s$. On examining it together with GPT-6 Astra (occasionally switched to a less advanced model, such as
GPT-5.6 Sol) and Claude Opus 5.5, the author found that the potential, the log-determinant, and Deninger's theorem
could all be bypassed in favor of the physical core of the calculation, namely, the minimum-norm solution of the
underdetermined tangent-space propagation problem. This solution directly gives the cumulative distribution as the
normalized trace of a one-step response matrix, the projector onto the decaying directions (Part~1,
Sec.~\ref{sec:part1}), obtained by propagating the decaying component of the source forward in time and the growing
component backward in time. Deninger's log-determinant encodes the same split, since the growing directions, which
must be propagated backward, are those that contribute their exponents. Indeed, Part~1, together with the differentiation above, gives an
elementary derivation of Deninger's theorem in this setting.

In this picture, each element of the original derivation acquires a physical meaning. The
regularizer selects the minimum-norm solution as it is taken to zero, and the $2\times2$ block system is the
forward--backward system of the corresponding least-squares problem (Part~2, Sec.~\ref{sec:dyn}). The present derivation also replaces the Gaussian-conditioning
argument and the matrix Dyson equation with a cavity calculation applied jointly to the network and the
forward--backward system (Part~3, Sec.~\ref{sec:part3}). In the cavity calculation, the couplings of the added neuron
are independent of the reservoir by construction, so that the dependence of the gains on the couplings is handled by
the order-of-magnitude estimates of Part~3, and the network dynamics and the forward--backward system reduce together to a single-site problem,
as in dynamical mean-field theory. Finally, the present derivation treats the Euler discretization with a finite time
step $\delta$. In the continuous-time limit $\delta\to0$, the resulting single-site equations coincide with those of
the original derivation.

\paragraph{Code, figures, and text}
Under the direction of the author, Claude Opus 5.5 wrote all code, both for the numerical solution of the theory and
for the network simulations, generated all figures, and checked the steps of the derivation numerically. Claude Opus
5.5 was also used to revise the text and to write initial drafts of some technical passages.

\clearpage
\bibliographystyle{unsrtnat}
\bibliography{refs}

\clearpage
\begin{appendices}
\numberwithin{equation}{section}

\section{Lyapunov spectrum at the trivial fixed point}
\label{app:tfp}

This appendix gives the details of the Lyapunov spectrum at the trivial fixed point (rather than along a chaotic
trajectory), summarized in Sec.~\ref{sec:limits}. We first obtain the
Lyapunov spectrum directly from the circular law of eigenvalues of $J$ (Appendix~\ref{app:tfp-exact}). We then show that the single-site
theory derived in this paper recovers this Lyapunov spectrum (Appendices~\ref{app:tfp-site}
and~\ref{app:tfp-gauss}).

\subsection{Lyapunov spectrum at the trivial fixed point from the circular law}
\label{app:tfp-exact}

\begin{figure}[t]
    \centering
    \includegraphics[width=\textwidth]{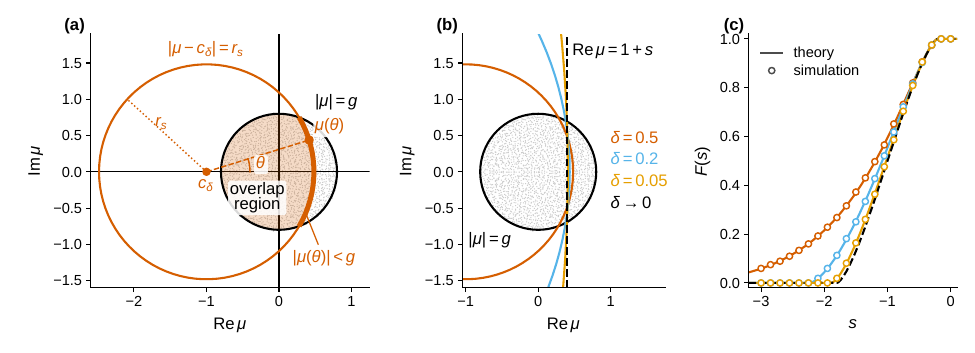}
    \caption{Trivial fixed point, $g=0.8$.
    \textbf{(a)} Eigenvalues $\mu_i$ of $J$ for one network with $N=2000$ (dots), the circle $|\mu|=g$, and the circle
    $|\mu-c_\delta|=r_s$ for $\delta=0.5$ and $s=-0.6$. Shading marks the overlap region of the
    disks $|\mu|<g$ and $|\mu-c_\delta|<r_s$, Eq.~\eqref{app:Fexact}. The point $\mu(\theta)=c_\delta+r_se^{i\theta}$
    moves along the circle $|\mu-c_\delta|=r_s$ as $\theta$ varies. The thick arc, where $|\mu(\theta)|<g$, is the band
    of frequencies $|\hat L(\theta)|^2<\gamma_s$ of the single-site solution (Appendix~\ref{app:tfp-gauss}).
    \textbf{(b)} Circles $|\mu-c_\delta|=r_s$ for $\delta=0.5$, $0.2$, and $0.05$ at $s=-0.6$, with the legend in the panel,
    and the eigenvalues and the circle $|\mu|=g$ as in (a). As $\delta\to0$, the circle
    $|\mu-c_\delta|=r_s$ approaches the line $\re\mu=1+s$ (dashed).
    \textbf{(c)} Cumulative distribution $F(s)$ for the values of $\delta$ in (b), with the legend of (b). Lines show the
    single-site theory, Eq.~\eqref{app:Ffp}, which coincides with the circular-law result, Eq.~\eqref{app:Fexact}, and
    dots show the fraction of the exponents of Eq.~\eqref{app:lam} below $s$ for the network in (a). The dashed line
    shows the $\delta\to0$ limit, Eq.~\eqref{app:semicircle}.}
    \label{fig:tfp}
\end{figure}

For $g<1$, the fixed point $x_i(n)=0$ of the network is stable. At this fixed point, every gain is
$d_i(n)=\phi'(0)=1$, so the Jacobian is constant, $M(n)=(1-\delta)I_N+\delta J$, with eigenvalues $1-\delta+\delta\mu_i$,
where $\mu_i$ are the eigenvalues of $J$. For a constant Jacobian, the $m$th roots of the singular values of the product
$M(m)\cdots M(1)$ converge, as $m\to\infty$, to the moduli of the eigenvalues of $M(n)$, so that the Lyapunov exponents
are
\begin{equation}
    \lambda_i=\frac1\delta\log|1-\delta+\delta\mu_i|.
    \label{app:lam}
\end{equation}
The condition $\lambda_i<s$ is equivalent to
\begin{equation}
    |\mu_i-c_\delta|<r_s,
\end{equation}
with
\begin{align}
    c_\delta&=1-\frac1\delta,\\
    r_s&=\frac{e^{\delta s}}{\delta}.
\end{align}
As $N\to\infty$, the eigenvalues $\mu_i$ are uniform on the disk $|\mu|<g$, by the circular law~\cite{ginibre1965statistical,girko1985circular,sommers1988spectrum}. The
fraction of exponents below $s$ is therefore the fraction of the disk $|\mu|<g$ that lies inside the disk
$|\mu-c_\delta|<r_s$. We call the region in which the two disks overlap the overlap region (Fig.~\ref{fig:tfp}a). Thus,
\begin{equation}
    F(s)=\frac{1}{\pi g^2}\,\mathrm{Area}\left\{\mu\in\mathbb C:|\mu|<g\ \text{and}\ |\mu-c_\delta|<r_s\right\}.
    \label{app:Fexact}
\end{equation}

As $\delta\to0$, Eq.~\eqref{app:lam} gives
\begin{equation}
    \lambda_i\to-1+\re\mu_i,
\end{equation}
the real parts of the eigenvalues of the continuous-time Jacobian $-I_N+J$. Correspondingly, the center $c_\delta$
of the circle $|\mu-c_\delta|=r_s$ moves to $-\infty$, while the rightmost point of this circle,
$\mu=c_\delta+r_s=1+s+\order{\delta}$, converges to $1+s$, so that, near the disk $|\mu|<g$, the disk
$|\mu-c_\delta|<r_s$ becomes the half-plane $\re\mu<1+s$ (Fig.~\ref{fig:tfp}b). Thus, as $\delta\to0$, $F(s)$ is the
fraction of the disk $|\mu|<g$ with $\re\mu<1+s$,
\begin{equation}
    F(s)=\frac{2}{\pi g^2}\int_{-g}^{1+s}\sqrt{g^2-x^2}\,dx,\qquad|1+s|<g,
    \label{app:semicircle}
\end{equation}
where the integrand, the projection of the uniform distribution on the disk $|\mu|<g$ onto the real axis, is the
semicircle law noted by \citet{engelken2023lyapunov}.

At $\delta=1$, where the network has no leak, Eq.~\eqref{app:lam} gives
\begin{equation}
    \lambda_i=\log|\mu_i|,
\end{equation}
the logarithms of the moduli of the eigenvalues of $J$. Correspondingly, $c_\delta=0$ and $r_s=e^s$, so that the disks
$|\mu|<g$ and $|\mu|<e^s$ are concentric. Thus, at $\delta=1$, $F(s)$ is the fraction of the disk $|\mu|<g$ with
$|\mu|<e^s$,
\begin{equation}
    F(s)=\frac{e^{2s}}{g^2},\qquad s<\log g,
    \label{app:delta1}
\end{equation}
as obtained by \citet{curato2013onset}.

\subsection{Lyapunov spectrum from the single-site theory}
\label{app:tfp-site}

We now show that the single-site theory recovers the Lyapunov spectrum of Eq.~\eqref{app:Fexact}. In the
single-site theory, $x_0(n)=0$ solves the mean-field theory, Eq.~\eqref{eq:dmft}, since $\phi(0)=0$. Thus, $d_0(n)=1$, $D_0=I$,
and the averages $\tavg{\cdot}_{x_0}$ are trivial. Equation~\eqref{eq:closed} becomes $A(n,n')=\gamma_sR^{vu}_{00}(n,n')$ and
$P(n,n')=R^{uv}_{00}(n,n')$, and we look for a solution in which all matrices in Eqs.~\eqref{eq:Rvu00}--\eqref{eq:closed} have entries
that depend only on $n-n'$. For such a matrix $Z$, we use the Fourier representation
\begin{equation}
    Z(n,n')=\int_{-\pi}^{\pi}\frac{d\theta}{2\pi}\,\hat Z(\theta)\,e^{i\theta(n-n')}.
\end{equation}
Products of such matrices become products of the functions $\hat Z(\theta)$, and inverses become reciprocals. For $L$
and $L^T$,
\begin{align}
    \hat L(\theta)&=e^{i\theta}-\alpha_s,\\
    \widehat{L^T}(\theta)&=e^{-i\theta}-\alpha_s,
\end{align}
so that $\hat L(\theta)\,\widehat{L^T}(\theta)=|\hat L(\theta)|^2$. Since the kernels are symmetric (Sec.~\ref{sec:part3}), $\hat A(\theta)$ and $\hat P(\theta)$ are real. We define
\begin{align}
    X(\theta)&=\eta+\hat A(\theta),\\
    Y(\theta)&=\eta+\gamma_s\hat P(\theta),\\
    Q(\theta)&=X(\theta)\,Y(\theta)+|\hat L(\theta)|^2.
\end{align}
Equations~\eqref{eq:Rvu00}--\eqref{eq:Rvv00} then give
\begin{align}
    \hat R^{vu}_{00}(\theta)&=\frac{X(\theta)}{Q(\theta)},\\
    \hat R^{uv}_{00}(\theta)&=\frac{Y(\theta)}{Q(\theta)},\\
    \hat R^{vv}_{00}(\theta)&=\frac{e^{-i\theta}-\alpha_s}{Q(\theta)}.
\end{align}
The self-consistency conditions, $\hat A(\theta)=\gamma_s\hat R^{vu}_{00}(\theta)$ and
$\gamma_s\hat P(\theta)=\gamma_s\hat R^{uv}_{00}(\theta)$, read
\begin{align}
    X(\theta)\left[1-\frac{\gamma_s}{Q(\theta)}\right]&=\eta,\\
    Y(\theta)\left[1-\frac{\gamma_s}{Q(\theta)}\right]&=\eta.
\end{align}
The right sides are $\eta>0$, so the factor $1-\gamma_s/Q(\theta)$ on the left sides cannot vanish, and dividing the
two equations gives
$X(\theta)=Y(\theta)$, so that $Q(\theta)=X(\theta)^2+|\hat L(\theta)|^2$. Since $A(n,n')$ is positive semidefinite,
as the response $R^{vu}=(\eta I+\eta^{-1}K^TK)^{-1}$ of Eq.~\eqref{eq:B} is positive definite,
$X(\theta)=\eta+\hat A(\theta)\geq\eta$. Thus, $X(\theta)$ is the positive root
of
\begin{equation}
    X(\theta)\left[1-\frac{\gamma_s}{X(\theta)^2+|\hat L(\theta)|^2}\right]=\eta.
\end{equation}
As $\eta\to0^+$, this root behaves differently inside and outside the band of frequencies where
$|\hat L(\theta)|^2<\gamma_s$,
\begin{align}
    X(\theta)^2&\to\begin{cases}\gamma_s-|\hat L(\theta)|^2,&|\hat L(\theta)|^2<\gamma_s,\\0,&|\hat L(\theta)|^2>\gamma_s,\end{cases}\\
    Q(\theta)&\to\begin{cases}\gamma_s,&|\hat L(\theta)|^2<\gamma_s,\\|\hat L(\theta)|^2,&|\hat L(\theta)|^2>\gamma_s.\end{cases}
\end{align}
The readout, Eq.~\eqref{eq:Fsite}, is
\begin{equation}
    F(s)=\lim_{\eta\to0^+}\int_{-\pi}^{\pi}\frac{d\theta}{2\pi}\,e^{i\theta}\,\hat R^{vv}_{00}(\theta)
    =\lim_{\eta\to0^+}\int_{-\pi}^{\pi}\frac{d\theta}{2\pi}\,\frac{1-\alpha_se^{i\theta}}{Q(\theta)}.
\end{equation}
Outside the band, $(1-\alpha_se^{i\theta})/|\hat L(\theta)|^2=1/(1-\alpha_se^{-i\theta})$. Dropping the parts of the
integrand that are odd in $\theta$, we obtain
\begin{equation}
    F(s)=\int_{|\hat L(\theta)|^2<\gamma_s}\frac{d\theta}{2\pi}\,\frac{1-\alpha_s\cos\theta}{\gamma_s}
    +\int_{|\hat L(\theta)|^2>\gamma_s}\frac{d\theta}{2\pi}\,\re\frac{1}{1-\alpha_se^{-i\theta}}.
    \label{app:Ffp}
\end{equation}

\subsubsection{Electrostatic interpretation}
\label{app:tfp-gauss}

To show that Eq.~\eqref{app:Ffp} coincides with Eq.~\eqref{app:Fexact}, we interpret Eq.~\eqref{app:Ffp} in
two-dimensional electrostatics, as is common for the eigenvalue densities of non-Hermitian random
matrices~\cite{sommers1988spectrum}. We first parametrize the circle
$|\mu-c_\delta|=r_s$ by an angle around $c_\delta$,
\begin{equation}
    \mu(\theta)=c_\delta+r_se^{i\theta}.
\end{equation}
This angle is the frequency $\theta$ of Eq.~\eqref{app:Ffp}, because $c_\delta=-\alpha_s/\beta_s$ and $r_s=1/\beta_s$, by Eqs.~\eqref{eq:alpha} and~\eqref{eq:beta}, so
that
\begin{equation}
    \beta_s\,\mu(\theta)=e^{i\theta}-\alpha_s=\hat L(\theta),
\end{equation}
and therefore $|\hat L(\theta)|^2=\beta_s^2|\mu(\theta)|^2$. Since $\gamma_s=\beta_s^2g^2$, the band
$|\hat L(\theta)|^2<\gamma_s$ consists of the angles at which $|\mu(\theta)|<g$ (thick arc in Fig.~\ref{fig:tfp}a).
In terms of $\mu(\theta)$, the two integrands of Eq.~\eqref{app:Ffp} are
\begin{alignat}{2}
    \frac{1-\alpha_s\cos\theta}{\gamma_s}&=\frac{\im\big[\bar\mu(\theta)\,\mu'(\theta)\big]}{g^2},&\qquad&|\mu(\theta)|<g,\\
    \re\frac{1}{1-\alpha_se^{-i\theta}}&=\im\frac{\mu'(\theta)}{\mu(\theta)},&&|\mu(\theta)|>g.
\end{alignat}
Therefore,
\begin{equation}
    F(s)=\int_{-\pi}^{\pi}\frac{d\theta}{2\pi}\,\im\big[\mathcal G(\mu(\theta))\,\mu'(\theta)\big]
    =\frac{1}{2\pi}\,\im\oint_{|\mu-c_\delta|=r_s}\mathcal G(\mu)\,d\mu,
    \label{app:FG}
\end{equation}
with
\begin{equation}
    \mathcal G(\mu)=\begin{cases}\bar\mu/g^2,&|\mu|\leq g,\\1/\mu,&|\mu|>g.\end{cases}
\end{equation}
The real and imaginary parts of $\mathcal G(\mu)$ are the components $\mathcal E_x(\mu)$ and $-\mathcal E_y(\mu)$ of the
electric field in a two-dimensional electrostatics problem sourced by the eigenvalues, with the real and imaginary parts
of $\mu=x+iy$ as the coordinates $x$ and $y$. In this problem, a unit point charge at $\nu$ produces at $\mu$ the field
\begin{equation}
    \mathcal E_x(\mu)-i\mathcal E_y(\mu)=\frac{1}{\mu-\nu}.
\end{equation}
Spreading a unit total charge uniformly over the disk $|\nu|<g$, as the eigenvalues are spread for $N\to\infty$, and
integrating this field over the disk gives
\begin{equation}
    \mathcal E_x(\mu)-i\mathcal E_y(\mu)=\frac{1}{\pi g^2}\int_{|\nu|<g}\frac{d^2\nu}{\mu-\nu}=\mathcal G(\mu).
\end{equation}
With $d\mu=dx+i\,dy$, $\im[\mathcal G(\mu)\,d\mu]=\mathcal E_x(\mu)\,dy-\mathcal E_y(\mu)\,dx$ is the flux of the electric field through the line
element $d\mu$. By Eq.~\eqref{app:FG}, $2\pi F(s)$ is therefore the flux of the electric field through the circle
$|\mu-c_\delta|=r_s$. By Gauss's law, this flux is $2\pi$ times the charge enclosed by the circle $|\mu-c_\delta|=r_s$,
since a point charge $q$ contributes a flux $2\pi q$ if the point charge lies inside the circle
$|\mu-c_\delta|=r_s$ and $0$ otherwise. The enclosed charge is the fraction of the disk $|\mu|<g$ that lies inside the
disk $|\mu-c_\delta|<r_s$, which is Eq.~\eqref{app:Fexact}. Thus, the single-site theory recovers the Lyapunov spectrum of Eq.~\eqref{app:Fexact}
for every $\delta$ (Fig.~\ref{fig:tfp}c), and therefore also the semicircle law as $\delta\to0$.

\section{Maximum Lyapunov exponent}
\label{app:maxexp}

This appendix gives the details of the maximum Lyapunov exponent, summarized in Sec.~\ref{sec:limits}. The maximum
Lyapunov exponent $\lambda_1$ is the rate at which two nearby trajectories of the same network separate, for a generic
displacement between them. At large $N$, this rate follows from dynamical mean-field theory applied to two copies of
the network, as in \citet{crisanti2018path} for $\delta\to0$ and \citet{molgedey1992suppressing} for
$\delta=1$. We first carry out this calculation at finite $\delta$. The calculation shows that $\lambda_1$ is
determined by a linear operator $\mathcal T_s$ (Appendix~\ref{app:replica}). We then evaluate $\lambda_1$ in limiting
cases (Appendix~\ref{app:limits}). Finally, and most importantly, we show that the single-site theory derived in this
paper recovers this maximum Lyapunov exponent (Appendix~\ref{app:edge}).

Throughout, we use the single-site autocovariance $C^x(\tau)$ (Sec.~\ref{sec:dimension}) and the gain
autocorrelation $C^d(\tau)$
(Sec.~\ref{sec:limits}) at the physical lag $\tau=(n-n')\delta$.

\subsection{Maximum Lyapunov exponent from two copies of the network}
\label{app:replica}

Consider two copies of the network, $x^{(1)}_i(n)$ and $x^{(2)}_i(n)$, with the same couplings $J_{ij}$ and different
initial conditions. At large $N$, the squared distance per neuron between the two copies,
$N^{-1}\sum_{i=1}^N[x^{(1)}_i(n)-x^{(2)}_i(n)]^2$, is self-averaging. For nearby copies, this squared distance grows at
the rate $2\lambda_1$, which we now compute following \citet{engelken2022input}.

The cavity argument of Sec.~\ref{sec:part3}, applied to both copies, gives, for neuron $0$,
\begin{align}
    x^{(1)}_0(n+1)&=(1-\delta)\,x^{(1)}_0(n)+\delta\,\xi^{(1)}_0(n),\\
    x^{(2)}_0(n+1)&=(1-\delta)\,x^{(2)}_0(n)+\delta\,\xi^{(2)}_0(n),
\end{align}
where the neuronal cavity fields $\xi^{(1)}_0(n)$ and $\xi^{(2)}_0(n)$ are jointly Gaussian. We write $\tavg{\cdot}_{x^{(1)}_0,x^{(2)}_0}$
for the average over this joint process. Each copy separately obeys the mean-field theory of Eq.~\eqref{eq:dmft}, and the
cross covariance of the two neuronal cavity fields is
\begin{equation}
    \tavg{\xi^{(1)}_0(n)\,\xi^{(2)}_0(n')}_{x^{(1)}_0,x^{(2)}_0}=g^2\tavg{\phi(x^{(1)}_0(n))\,\phi(x^{(2)}_0(n'))}_{x^{(1)}_0,x^{(2)}_0}.
\end{equation}
Applying the update of the first copy to the first time argument and the update of the second copy to the second time
argument, the cross covariance $C^{12}(n,n')=\tavg{x^{(1)}_0(n)\,x^{(2)}_0(n')}_{x^{(1)}_0,x^{(2)}_0}$ obeys
\begin{multline}
    C^{12}(n+1,n'+1)-(1-\delta)\big[C^{12}(n,n'+1)+C^{12}(n+1,n')\big]+(1-\delta)^2C^{12}(n,n')\\
    =\delta^2g^2\tavg{\phi(x^{(1)}_0(n))\,\phi(x^{(2)}_0(n'))}_{x^{(1)}_0,x^{(2)}_0}.
\end{multline}
Identical copies give the solution $C^{12}(n,n')=C^x(\tau)$. For nearby copies, we write
$C^{12}(n,n')=C^x(\tau)+\psi(n,n')$, so that the mean squared distance between the two copies of neuron $0$, which
equals the squared distance per neuron, is
\begin{equation}
    \tavg{\big[x^{(1)}_0(n)-x^{(2)}_0(n)\big]^2}_{x^{(1)}_0,x^{(2)}_0}=-2\psi(n,n).
\end{equation}
To first order in $\psi(n,n')$, only the covariance of the jointly Gaussian variables $x^{(1)}_0(n)$ and
$x^{(2)}_0(n')$ changes, so, by Price's theorem~\cite{price1958useful}, the right side of the equation for $C^{12}(n,n')$
changes by $\delta^2g^2C^d(\tau)\,\psi(n,n')$. Thus, $\psi(n,n')$ obeys the linear equation
\begin{equation}
    \psi(n+1,n'+1)-(1-\delta)\big[\psi(n,n'+1)+\psi(n+1,n')\big]+(1-\delta)^2\psi(n,n')
    =\delta^2g^2C^d(\tau)\,\psi(n,n').
\end{equation}

We look for solutions that grow exponentially at a rate $s$, $\psi(n,n')=e^{\delta s(n+n')}\psi(\tau)$, for which the
squared distance $-2\psi(n,n)$ grows as $e^{2\delta sn}$, so that the distance grows at the rate $s$. Inserting this form, dividing by
$e^{\delta s(n+n'+2)}$, and using $\alpha_s=(1-\delta)e^{-\delta s}$ and $\gamma_s=\delta^2g^2e^{-2\delta s}$ gives
\begin{equation}
    (\mathcal T_s\psi)(\tau)=0,
    \label{app:zeromode}
\end{equation}
where the linear operator $\mathcal T_s$ acts on functions $\psi(\tau)$ of the lag as
\begin{equation}
    (\mathcal T_s\psi)(\tau)=\big(1+\alpha_s^2\big)\psi(\tau)-\alpha_s\big[\psi(\tau+\delta)+\psi(\tau-\delta)\big]
    -\gamma_s\,C^d(\tau)\,\psi(\tau).
    \label{app:T}
\end{equation}

A small initial separation gives a perturbation $\psi(n,n')$ that is bounded in $\tau$. Thus, the distance can grow at
the rate $s$ only if Eq.~\eqref{app:zeromode} has a solution $\psi(\tau)$ that does not grow at large $|\tau|$. Such a solution exists exactly when $0$ lies in the spectrum of $\mathcal T_s$ acting on bounded functions of $\tau$. The distance between the copies grows at the largest such rate. Thus,
\begin{equation}
    \lambda_1=s_*,
\end{equation}
where $s_*$ is the largest value of $s$ at which $0$ lies in the spectrum of $\mathcal T_s$.

\subsubsection{Limiting cases}
\label{app:limits}

\begin{figure}[t]
    \centering
    \includegraphics[width=\textwidth]{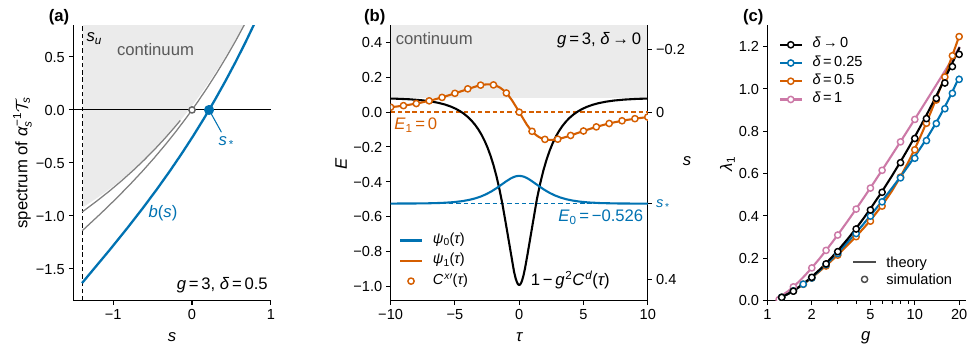}
    \caption{Maximum Lyapunov exponent from the operator $\mathcal T_s$, Eq.~\eqref{app:T}.
    \textbf{(a)} Low-lying spectrum of $\alpha_s^{-1}\mathcal T_s$, whose eigenvalues have the same signs as those of
    $\mathcal T_s$, as a function of $s$ for
    $g=3$ and $\delta=0.5$. The thick line shows the lowest value $b(s)$ of the spectrum, and the thin lines show excited
    eigenvalues below the continuum (shading). The dashed line
    marks $s_u=\delta^{-1}\log(1-\delta)$, and the filled circle marks the zero crossing $s_*$ of $b(s)$. The first excited
    eigenvalue crosses zero close to $s=0$ (open circle), the finite-$\delta$ counterpart of the zero mode
    $C^{x\prime}(\tau)$ in (b).
    \textbf{(b)} Schr\"odinger problem of the limit $\delta\to0$, Eq.~\eqref{app:schrodinger}, for $g=3$. The line
    labeled $1-g^2C^d(\tau)$ shows the potential of $H$, Eq.~\eqref{app:H}, and the shading shows the continuum of $H$,
    above $1-g^2\tavg{d_0(n)}_{x_0}^2$. The bound states $\psi_0(\tau)$ and $\psi_1(\tau)$ of $H$, with the legend in the
    panel, are drawn at their energies $E_0$ and $E_1$ (dashed lines), and dots show $C^{x\prime}(\tau)$, which is an eigenfunction of $H$ with
    energy zero~\cite{sompolinsky1988chaos}, normalized like $\psi_1(\tau)$. The right axis gives the exponent $s=-1+\sqrt{1-E}$ that
    corresponds to the energy $E$, so that the ground state gives $s_*=-1+\sqrt{1-E_0}$, and the zero mode
    $C^{x\prime}(\tau)$, with $E_1=0$, corresponds to $s=0$.
    \textbf{(c)} Maximum Lyapunov exponent $\lambda_1$ as a function of
    $g$ for $\delta\to0$, $0.25$, $0.5$, and $1$, with the legend in the panel. Lines show $s_*$ from $\mathcal T_s$, and
    dots show simulations with $N=4096$ (larger networks near the onset of chaos for $\delta<1$;
    Appendix~\ref{app:num-sim}).}
    \label{fig:maxexp}
\end{figure}

We evaluate $s_*$ in two limiting cases. First, as $\delta\to0$, we expand
$\psi(\tau\pm\delta)=\psi(\tau)\pm\delta\psi'(\tau)+\frac12\delta^2\psi''(\tau)+\order{\delta^3}$,
$\alpha_s=1-\delta(1+s)+\order{\delta^2}$, and $\gamma_s=\delta^2g^2+\order{\delta^3}$. The terms of order $1$
and $\delta$ in Eq.~\eqref{app:T} cancel, since $1+\alpha_s^2-2\alpha_s=(1-\alpha_s)^2=\delta^2(1+s)^2+\order{\delta^3}$,
and Eq.~\eqref{app:T} becomes
\begin{equation}
    (\mathcal T_s\psi)(\tau)=\delta^2\big[(H\psi)(\tau)-E\,\psi(\tau)\big]+\order{\delta^3},
\end{equation}
with
\begin{align}
    H&=-\frac{d^2}{d\tau^2}+1-g^2C^d(\tau),\label{app:H}\\
    E&=1-(1+s)^2.
\end{align}
Equation~\eqref{app:zeromode} thus becomes the Schr\"odinger equation
\begin{equation}
    (H\psi)(\tau)=E\,\psi(\tau).
    \label{app:schrodinger}
\end{equation}
The operator $H$ does not depend on $s$, and all of the dependence on $s$ is in the energy $E$. Thus, $0$ lies in
the spectrum of $\mathcal T_s$ when $E$ lies in the spectrum of $H$. As $s$ increases from $-1$, $E$ decreases from
$1$. The largest $s$ at which $E$ lies in the spectrum of $H$ is therefore reached when $E$ reaches the lowest value of
this spectrum, the ground-state energy $E_0$, which lies below $1$, since the potential $1-g^2C^d(\tau)$ is less than
$1$ everywhere (Fig.~\ref{fig:maxexp}b). Thus,
\begin{equation}
    s_*=-1+\sqrt{1-E_0},
\end{equation}
which is the known result~\cite{sompolinsky1988chaos}.

Second, at $\delta=1$, where the network has no leak, $\alpha_s=0$, and
$\mathcal T_s$ multiplies by $1-g^2e^{-2s}C^d(\tau)$. The eigenfunctions of $\mathcal T_s$ are then concentrated at
single lags $\tau$, with eigenvalues $1-g^2e^{-2s}C^d(\tau)$. Since $C^d(\tau)$ is largest at $\tau=0$, the largest
$s$ at which one of these eigenvalues vanishes is
\begin{equation}
    s_*=\log\big(g\sqrt{C^d(0)}\big),
\end{equation}
with $C^d(0)=\tavg{d_0(n)^2}_{x_0}$, which is the known result~\cite{molgedey1992suppressing}.

\subsection{Maximum Lyapunov exponent from the single-site theory}
\label{app:edge}

We now show that the single-site theory recovers $\lambda_1=s_*$, as the upper edge of the Lyapunov spectrum, above
which $F(s)=1$. Above $\lambda_1$, every tangent direction decays under the shifted dynamics, so that
the minimum-norm solution of Part~1 (Sec.~\ref{sec:part1}) is the causal solution, which propagates the whole source
forward. In the single-site theory, the analog of the causal solution arises when the kernels $A(n,n')$ and $P(n,n')$
vanish as $\eta\to0^+$, so that the forward field of neuron $0$ follows the uncoupled shifted dynamics. We first show
that vanishing kernels give $F(s)=1$, and then determine for which $s$ Eq.~\eqref{eq:closed} has such a solution.

When the kernels vanish as $\eta\to0^+$, so do $\eta I+A$ and $\eta I+\gamma_s\,D_0PD_0$. The inverses in the last
terms of Eqs.~\eqref{eq:Rvu00}--\eqref{eq:Ruv00} then diverge and dominate, so that, to leading order,
\begin{align}
    R^{vu}_{00}&\approx L^{-1}\big(\eta I+A\big)L^{-T},\\
    R^{uv}_{00}&\approx L^{-T}\big(\eta I+\gamma_s\,D_0PD_0\big)L^{-1}.
\end{align}
Equation~\eqref{eq:Rvv00} then gives $R^{vv}_{00}\to L^{-1}$. For $s>s_u$, where $s_u=\delta^{-1}\log(1-\delta)$ is the exponent at which $\alpha_s=1$, $L^{-1}$ propagates the
uncoupled shifted dynamics forward in time,
\begin{equation}
    L^{-1}(n,n')=\begin{cases}\alpha_s^{\,n-n'-1},&n>n',\\0,&n\leq n'.\end{cases}
\end{equation}
In particular, $R^{vv}_{00}(k+1,k)\to L^{-1}(k+1,k)=1$, so that $F(s)=1$.

With these leading-order responses, Eq.~\eqref{eq:closed} gives kernels that are linear in $\eta$ and in the kernels
themselves, so that the kernels $A(n,n')$ and $P(n,n')$ are proportional to $\eta$. We therefore write
\begin{align}
    A(n,n')&=\eta\,A_1(n,n'),\\
    \gamma_sP(n,n')&=\eta\,P_1(n,n').
\end{align}
For a deterministic matrix $Z$, $\tavg{D_0ZD_0}_{x_0}$ has entries $C^d(\tau)\,Z(n,n')$, which we write as
$C^d\circ Z$. Equation~\eqref{eq:closed}, at leading order in $\eta$, becomes
\begin{align}
    A_1&=\gamma_s\,C^d\circ\Sigma,\\
    P_1&=\gamma_s\,L^{-T}\big(I+C^d\circ P_1\big)L^{-1},
\end{align}
where $\Sigma=L^{-1}(I+A_1)L^{-T}$. Eliminating $A_1$, these read
\begin{align}
    L\,\Sigma\,L^T-\gamma_s\,C^d\circ\Sigma&=I,\\
    L^TP_1L-\gamma_s\,C^d\circ P_1&=\gamma_s I.
\end{align}
For a matrix $Z$ whose entries $Z(\tau)$ depend only on the lag, both $(LZL^T)(n,n')$ and $(L^TZL)(n,n')$ are equal to
$(1+\alpha_s^2)Z(\tau)-\alpha_s[Z(\tau+\delta)+Z(\tau-\delta)]$. In the stationary state, both equations therefore apply
the operator $\mathcal T_s$ of Eq.~\eqref{app:T}, to $\Sigma(\tau)$ and to $P_1(\tau)$, with sources that differ only by
the factor $\gamma_s$. Thus, $P_1(\tau)=\gamma_s\Sigma(\tau)$, and both kernels, $A_1(\tau)=\gamma_sC^d(\tau)\,\Sigma(\tau)$
and $P_1(\tau)$, are determined by the single equation
\begin{equation}
    (\mathcal T_s\Sigma)(\tau)=\begin{cases}1,&\tau=0,\\0,&\tau\neq0.\end{cases}
\end{equation}

The same operator $\mathcal T_s$ as in the calculation with two copies of the network thus decides whether the vanishing-kernel solution
exists. For $s>s_*$, which exceeds $s_u$, $0$ does not lie in the spectrum of $\mathcal T_s$, so that $\mathcal T_s$ is invertible, and the equation for $\Sigma(\tau)$ has a
square-summable solution. As $s$ decreases to $s_*$, $0$ enters the spectrum of $\mathcal T_s$, and $\Sigma(\tau)$ diverges. Thus, the solution with vanishing kernels, and with it $F(s)=1$,
exists for $s>s_*$ but cannot be continued below $s_*$. Below $s_*$, the numerical solution of the single-site theory
has kernels of order one as $\eta\to0^+$ and gives $F(s)<1$ (Fig.~\ref{fig:spectra}). Thus, the upper edge of the Lyapunov
spectrum in the single-site theory is $s_*$, so that the single-site theory recovers $\lambda_1=s_*$ for every $\delta$,
and therefore also the results of \citet{sompolinsky1988chaos} and \citet{molgedey1992suppressing} as $\delta\to0$ and at $\delta=1$,
respectively.

\section{Participation-ratio dimension at finite time step}
\label{app:pr}

This appendix gives the details of the participation-ratio dimension at finite time step $\delta$, summarized in
Sec.~\ref{sec:dimension}. The two-site cavity derivation~\cite{clark2023dimension} carries over step by step to discrete
time, with three differences. First, the single-site dynamics are those of the Euler map,
Eq.~\eqref{eq:x0lim}, so that the response $S^x(\tau)$ of $x_0(n)$ to $\xi_0(n')$, with $\tau=(n-n')\delta$, is
$\delta(1-\delta)^{n-n'-1}$ for $n>n'$ and zero otherwise. Second, frequencies lie on the Brillouin zone $\theta\in(-\pi,\pi]$ rather than on the real line. Third, we
use the Fourier transform
\begin{equation}
    \hat f(\theta)=\sum_{n\in\mathbb Z}f(n\delta)\,e^{-i\theta n},
\end{equation}
for functions $f(\tau)$ of the lag, so that the factors $2\pi$ of \citet{clark2023dimension} are absent. In this
convention, the transform of the single-site response $S^x(\tau)$ is
\begin{equation}
    \hat S^x(\theta)=\frac{\delta}{e^{i\theta}-1+\delta}.
\end{equation}
With $\hat\psi^a(\theta_1,\theta_2)$ the transform in both lags of the four-point function $\psi^a(\tau_1,\tau_2)$, the two-site cavity derivation gives
\begin{align}
    \hat\psi^\phi(\theta_1,\theta_2)&=\left[\frac{1}{|1-z(\theta_1,\theta_2)|^2}-1\right]
        \hat C^\phi(\theta_1)\,\hat C^\phi(\theta_2),\label{eq:psiphi}\\
    \hat\psi^x(\theta_1,\theta_2)&=\left[\frac{2-|z(\theta_1,\theta_2)|^2}{|1-z(\theta_1,\theta_2)|^2}-1\right]
        \hat C^x(\theta_1)\,\hat C^x(\theta_2),\label{eq:psix}
\end{align}
with
\begin{equation}
    z(\theta_1,\theta_2)=g^2\tavg{d_0(n)}_{x_0}^2\,\hat S^x(\theta_1)\,\hat S^x(\theta_2),
\end{equation}
where $\hat C^\phi(\theta)$ and $\hat C^x(\theta)$ are the transforms of the autocovariances $C^\phi(\tau)$ and
$C^x(\tau)$ given by the mean-field theory, Eq.~\eqref{eq:dmft}. Integrating over the Brillouin zone gives the value $\psi^a(0,0)$ that enters $\mathrm{PR}^a/N$ (Sec.~\ref{sec:dimension}),
\begin{equation}
    \psi^a(0,0)=\int_{-\pi}^{\pi}\frac{d\theta_1}{2\pi}\int_{-\pi}^{\pi}\frac{d\theta_2}{2\pi}\,
    \hat\psi^a(\theta_1,\theta_2).
    \label{eq:psi00}
\end{equation}
As $\delta\to0$, at fixed frequency $\omega=\theta/\delta$, $\hat S^x(\theta)\to1/(1+i\omega)$, and
Eqs.~\eqref{eq:psiphi}--\eqref{eq:psi00} reduce to the continuous-time result~\cite{clark2023dimension}. To
bring Eq.~\eqref{eq:psix} into the form given there, we used
$\hat C^x(\theta)=g^2|\hat S^x(\theta)|^2\hat C^\phi(\theta)$, from Eq.~\eqref{eq:x0lim}, and
$\tavg{x_0(n)\,\phi(x_0(n'))}_{x_0}=\tavg{d_0(n)}_{x_0}C^x(\tau)$, from Gaussian integration by parts.

\section{Numerical methods}
\label{app:numerics}

This appendix gives the details of the numerical solution of the single-site theory and of the simulations,
summarized in Sec.~\ref{sec:numerics}.

\subsection{Single-site theory}
\label{app:num-site}

We iterate Eq.~\eqref{eq:closed} on a periodic window of $m$ steps, covering $m\delta=32$ or $64$ units of time, long
enough that doubling the window changes the results negligibly, so that the periodic window approximates the infinite
window of Part~1 (Sec.~\ref{sec:part1}). On this window, $L$, $A$, and $P$ are circulant
and are thus diagonalized by the discrete Fourier transform. In Eqs.~\eqref{eq:Rvu00}--\eqref{eq:Rvv00}, however,
$D_0$ appears inside the inverses, and the average in Eq.~\eqref{eq:closed} is taken outside these inverses. Since
$D_0$ is not translation invariant for a single gain trajectory, each sampled gain trajectory requires the inverse of
an $m\times m$ matrix in the time domain. We sample sets of $1024$ to $4096$ gain trajectories from the Gaussian process of
Eq.~\eqref{eq:dmft} using a scrambled Sobol sequence, with an independent scramble for each set, and average the
results over one to eight sets.

Near $s_u=\delta^{-1}\log(1-\delta)$, where $\alpha_s=1$ and the uncoupled shifted dynamics decay slowly, the
kernels decay slowly, and a periodic window of fixed length misrepresents the infinite window. There, we replace the periodic
boundary conditions of the window by twisted ones, under which a sequence acquires a phase $e^{i\varphi}$ after $m$
steps, and average $F(s)$ over $\varphi$ with a quadrature of two to four points. The average over $\varphi$ approximates the infinite window for a gain trajectory repeated periodically.

We lower $\eta$ from $0.1\,\delta$ to $1.6\times10^{-4}\,\delta$ (to $1.6\times10^{-6}\,\delta$ at $g=1.25$) in
steps, using each solution as the starting point for the next, and read out Eq.~\eqref{eq:Fsite} at the smallest
value. At finite $\delta$, our code also adds the term $\eta\tavg{R^{vu}_{00}(k,k)}_{x_0}$ to Eq.~\eqref{eq:Fsite}.
This term is a remnant of the original derivation (\hyperref[sec:ai]{AI methodology}) and has a negligible
effect on the results, since, at the smallest $\eta$, it is below $10^{-4}$ for every $s$, $g$, and $\delta$. We compute $F(s)$ on a grid of values of $s$, interpolate monotonically between them, and set $F(s)=1$ above
$s_*$, computed from $\mathcal T_s$ (Appendix~\ref{app:replica}). For $\delta\to0$, we solve the single-site problem in continuous time, in which $L$ becomes $\delta\,(d/dt+1+s)$.
In continuous time, a source at time $t'$ makes $v_0(t)$ jump at $t=t'$, and Eq.~\eqref{eq:Fsite} reads out
$R^{vv}_{00}(t,t')$ just after this jump. We represent time on a grid with spacing $0.4$ for $g\leq5$ and $0.2$ at
$g=6$. For $g\geq8$, where the gains change on a time scale of about $1/g$, we solve at spacings $0.2$ and $0.125$ and
extrapolate to zero spacing, assuming that the error is proportional to the square of the spacing.

\subsection{Simulations}
\label{app:num-sim}

We discard an initial transient of at least $300$ units of time, evolve the tangent vectors for a further $200$ units
of time, and then average over $2000$ to $8000$ units of time, reorthonormalizing the tangent vectors every $0.5$ units
of time (every step at $\delta=1$). For $\delta\to0$, we use the fourth-order Runge--Kutta method with time step $0.1$
($0.05$ for $g\geq6$) and propagate the tangent vectors with the exact Jacobian of the Runge--Kutta step. In
Fig.~\ref{fig:spectra}, we compute the full Lyapunov spectrum. For most networks in Figs.~\ref{fig:dimension_entropy}
and~\ref{fig:maxexp}, we compute only the largest $256$ to $1024$ exponents, which include every exponent above
$s_{\mathrm{KY}}$ and therefore suffice for $D_{\mathrm{KY}}$ and $h_{\mathrm{KS}}$. Near the onset of chaos
($g\leq1.5$), only a few exponents are positive at $N=4096$, and, for $\delta<1$, we use larger networks and the
median over five networks. In Figs.~\ref{fig:dimension_entropy} and~\ref{fig:maxexp}, the simulations of the
Lyapunov exponents use $N=4096$ for $g\geq2$ and, for $g\leq1.5$, the median over five networks with $N=16384$ for
$\delta=0.25$ and $0.5$ and with $N=8192$ for $\delta\to0$. At $g=3$, they use the median over three or four networks
with $N=4096$ for $\delta=0.25$, $0.5$, and $1$. In Fig.~\ref{fig:maxexp}c, the simulations also include one network
with $N=16384$ at $g=1.75$ for $\delta=0.25$ and $0.5$, and the simulations for $\delta=1$ use $N=4096$ at every $g$.

For the participation ratios, we simulate two networks with $N=16384$ and average the participation ratios over
these two networks. For $\delta\to0$, we use the fourth-order Runge--Kutta method with time step $0.1$ ($0.05$ for $g\geq12$). In each
network, we sample the activity every $0.5$ units of time over $2\times10^4$ to
$8\times10^4$ units of time and split the samples into $8$ blocks. Finite sampling biases
$\sum_{i,j=1}^NC^a_{ij}(0)^2$ in Eq.~\eqref{eq:PRdef} upward, and thus $\mathrm{PR}^a$ downward. We remove this bias by
estimating $\sum_{i,j=1}^NC^a_{ij}(0)^2$ from products of covariances measured in different blocks.

\end{appendices}

\end{document}